\documentclass[sigconf,authorversion, manuscript,screen, nonacm]{acmart}

\usepackage{booktabs, graphicx, array, makecell, enumerate, placeins, comment, longtable, tabularray}
\usepackage{csvsimple}[13]
\usepackage[utf8]{inputenc}
\usepackage[russian, main=english]{babel}
\usepackage{subcaption}

\usepackage{soul}
\usepackage[dvipsnames]{xcolor}

\AtBeginDocument{%
  }

\setcopyright{acmlicensed}
\copyrightyear{2026}
\acmYear{2026}
\acmDOI{XXXXXXX.XXXXXXX}

\acmJournal{PACMHCI}
\acmYear{2026} 
\acmISBN{978-1-4503-XXXX-X/2018/06}

\begin{document}

\title[Empire Amplifier]{Empire Amplifier: Uncovering and Contesting the Prioritization of Colonial Content on Platforms Through Community-Informed Algorithmic Auditing\textsuperscript{*}}

\author{Nel Escher}
\email{nelescher@berkeley.edu}
\affiliation{%
  \institution{University of California, Berkeley}
  \city{Berkeley}
  \state{California}
  \country{USA}
  \postcode{48109}
}

\author{Bakyt Yrysov}
\email{yrysov_b@auca.kg}
\affiliation{%
  \institution{American University of Central Asia}
  \city{Bishkek}
  \state{Chui}
  \country{Kyrgyzstan}
}

\author{Ashley McDermott}
\email {ashmac@umich.edu}
\affiliation{%
  \institution{University of Michigan}
  \city{Ann Arbor}
  \state{Michigan}
  \country{USA}
  \postcode{48109}
}

\author{Daniel Chechelnitsky}
\email {dchechel@andrew.cmu.edu}
\affiliation{%
  \institution{Carnegie Mellon University}
  \city{Pittsburgh}
  \state{Pennsylvania}
  \country{USA}
  \postcode{15213}
}

\author{Hermela Berehan Benyam}
\email {hberehan@umich.edu}
\affiliation{%
  \institution{University of Michigan}
  \city{Ann Arbor}
  \state{Michigan}
  \country{USA}
  \postcode{48109}
}

\author{Nikola Banovic}
\email {nbanovic@umich.edu}
\affiliation{%
  \institution{University of Michigan}
  \city{Ann Arbor}
  \state{Michigan}
  \country{USA}
  \postcode{48109}
}

\renewcommand{\shortauthors}{Escher, et al.}

\begin{abstract}
  Though online platforms claim to amplify Indigenous voices, Indigenous communities are worried that these systems are instead eroding their language and culture. We conduct a community-informed algorithmic audit to explore whether online platforms sustain or endanger Indigenous cultural practice. First, we review ethnographic research pertaining to the cultural anxieties of a specific Indigenous community, as Indigenous peoples are not a monolith. We consider concerns from Kyrgyz communities who believe that platforms are expanding Russia's linguistic influence and threatening their language. Next, we construct and conduct an algorithmic audit in conversation with the community. Our audit investigates deep-seated fears among Kyrgyz caregivers that YouTube encourages their children to speak Russian instead of Kyrgyz, their heritage language. We measure how the YouTube recommendation algorithm prioritizes content across Indigenous and non-Indigenous languages for child users. Our results validate caregiver concerns, as we find that YouTube primarily recommends non-Kyrgyz content to Kyrgyz children, even when children signal clear preferences for Kyrgyz content. Thus, platform recommendations reinforce Kyrgyz children's offline uptake of colonial language ideologies. Finally, we evaluate strategies to align platform behavior with Indigenous values. We identify effective end-user practices for reducing the proportion of Russian-language YouTube recommendations, like cross-generational device sharing. Overall, our work uncovers how platforms can amplify colonial influence, rather than revitalizing Indigenous cultural heritage. We encourage researchers to consider how algorithmic systems can reimpose oppressive power structures that decolonial efforts have sought to dismantle.

\end{abstract}



\received{13 May 2025}

\maketitle

\begingroup
\renewcommand\thefootnote{*}
\footnotetext{Accepted to the 2026 ACM Conference on Computer-Supported Cooperative Work and Social Computing (CSCW '26).}
\endgroup

\section{Introduction}

Social media platforms claim that they amplify Indigenous voices and incorporate marginalized values~\cite{ytipd2024, meta2025, snap2025}. Scholars have supported this view, optimistically hailing platforms as a resource for preserving Indigenous cultural heritage~\cite{dale2022case, moshnikov2022use, wang2022minority}. They point to online language reclamation projects such as the Chickasaw Word of the Day account on Twitter~\cite{davis2018talking}, videos on TikTok promoting the heritage language and dress of minority language speakers in China~\cite{wang2022minority}, and Facebook archives dedicated to sharing historical records in endangered South Asian languages~\cite{dale2022case} among many others. Proponents assume that online platforms connect members of endangered language communities and expand their domains of use~\cite{davis2018talking}.

However, Indigenous communities are concerned that online platforms are instead eroding their language and culture~\cite{coffey2021, Shahid_2023, Nigatu_2024}. In contrast with academic enthusiasm for language revitalization content on social media~\cite{dale2022case, davis2018talking, moshnikov2022use, wang2022minority}, Indigenous people worry that these systems cause their children to prefer majority languages like English~\cite{nyabola2021, millan2024}. Some users believe that platforms afford less visibility to content from Indigenous creators than that produced by white creators~\cite{mayworm2024, schiff2022}, reaffirming colonial hierarchies that cast local cultures, relations, and people as inferior \cite{tlostanova2012, fanon2007}. Even if these impacts are not intended by developers, communities question whether their values are protected by Big Tech. These concerns have not been pursued in scholarly evaluations of algorithmic systems, which typically follow Western anxieties~\cite{urman2024, metaxa2021} (e.g., audits frequently use American categories, such as framing political bias as Republicans vs. Democrats~\cite{lutz2021}). Though journalists, scholars, and activists have collected anecdotal reports of Indigenous concerns, no empirical, rigorous investigation has yet explored whether popular systems are operating as feared.

In this study, we use community-informed algorithmic auditing to explore whether an online platform sustains or imperils Indigenous culture. We employ an empirical and decolonial approach, so that the evidence we gather about platforms serves Indigenous people \cite{tlostanova2012, torres2017}. First, we draw upon qualitative research to understand Indigenous community concerns, surfacing particular cultural aspects that are perceived to be under threat from online platforms. Then, we create our algorithmic audit research questions in conversation with the community. Algorithmic audits provide evidence about the operation and impacts of algorithmic curation \cite{metaxa2021}, so we tune our inquiry to examine content related to the threatened cultural aspects. Finally, we evaluate different strategies to contest platform algorithms and better align recommendations with community values. This approach yields knowledge that is immediately useful to the impacted community. As it is inappropriate to assume all Indigenous peoples share the same views~\cite{noe2024}, we build our inquiry from an in-depth analysis of a specific community. We consider the context of Kyrgyzstan, where the Indigenous Kyrgyz population has long contended with Russia's imperial power, because the community has expressed serious anxieties about the impact of platforms on their Indigenous language.

Our audit investigates widespread, deep-seated fears among Kyrgyz caregivers that YouTube encourages their children to speak Russian instead of Kyrgyz, their heritage language. Our team surfaced this cultural distress in a prior ethnographic study conducted in Bishkek, the capital of Kyrgyzstan \cite{mcdermott2026}. In the Soviet Union, Russian was a required subject in schools \cite{Kirkwood1990LanguagePI, grenoble2003language, fierman1991language} and fluency was vital for career advancement \cite{2000_Hirsch, martin2001, Slezkine_1994, smith2006non}. After Kyrgyzstan declared independence, government entities and language activists tried to promote the Kyrgyz language \cite{derbisheva2009, huskey1995politics, korth2005language, alybaeva2023}. Kyrgyz parents believe YouTube undermines such language revitalization efforts, since their children consume predominantly Russian-language content on the platform and mimic language heard online. To determine whether the platform steers Kyrgyz children away from Indigenous language videos and toward Russian-language content, we conduct a sock-puppet algorithmic audit~\cite{sandvig2014auditing} that measures how the YouTube recommendation algorithm prioritizes content across Indigenous and non-Indigenous languages. Since sock-puppets can signal different interests to an audited platform, this approach allows us to investigate videos recommended to children. We design a series of experiments to test the search and recommendation algorithms, because children used these to locate YouTube content during prior participant observation. These experiments record the algorithmic recommendations presented in response to popular search queries in Kyrgyzstan or encountered while browsing the platform. Research assistants from a university in Bishkek labeled the languages that appeared in each of the 10,741 unique videos gathered during the experiments.

Our auditing results validate caregiver concerns, as we find that YouTube primarily recommends non-Indigenous content to Indigenous Kyrgyz children---even when children signal clear preferences for Indigenous content. Evaluating equal numbers of search terms in Russian and Kyrgyz, we find that YouTube prioritizes Russian-language content over Kyrgyz-language content, and that this skew is even greater for terms related to children's interests. Surprisingly, a child who signals a strong interest in Kyrgyz-language content (i.e., by watching a sequence of Kyrgyz cartoons) is likely to see fewer Kyrgyz-language recommendations while browsing than new users who have yet to signal their interests. This behavior exists despite significant revitalization efforts to post Kyrgyz-language children's content on the platform\footnote{E.g., International aid projects (``\foreignlanguage{russian}{ЮНИСЕФ Кыргызстан}''/UNICEF Kyrgyzstan, https://www.youtube.com/@UNICEFKyrgyzstan), state-sponsored channels (``\foreignlanguage{russian}{Телеканал Баластан}''/TV Channel Balastan, https://www.youtube.com/\foreignlanguage{russian}{@ТелеканалБаластан}), and internationally popular creators (``D Billions \foreignlanguage{russian}{Кыргызча}'', https://www.youtube.com/@dbillions\_kyrgyz) post Kyrgyz-language cartoons and other content aimed at children.}. The algorithm did not alter its recommendations based on the composition of the linguistic environment; we obtained similar results in Bishkek, the mixed-language capital, and Kochkor, a rural village where Russian is rarely spoken. Our results indicate that the YouTube platform does not deliver desired content to users with Indigenous preferences, instead amplifying imperial languages. In response, we explore end-user strategies for contesting the algorithm and identify several that significantly reduce the proportion of Russian-language results. For instance, families can share a device, rather than give children their own phones. Thus, though the platform issues recommendations in response to current Kyrgyz technology practices that conflict with Indigenous values, this outcome should not be viewed as inevitable.

Claims for Indigenous self-determination must be recognized as extending to the digital sphere; current platforms insufficiently safeguard Indigenous cultural heritage. Though some scholars have charged Silicon Valley with extending American imperialism \cite{mcelroy2024silicon, little2021new}, we find that YouTube, an American platform, also projects Russia's influence, reproducing the linguistic hierarchies of the Russian imperial core in the digital spaces of its former colonies in Central Asia. Contrary to optimistic beliefs that online platforms will revitalize Indigenous culture, we find that these systems can amplify non-Indigenous material that reinforces atrophy of Indigenous practices offline. Alongside decolonial computing projects that envision new technologies made by or alongside peripheral populations~\cite{longdon2024, millan2024, arendttorp2022, muashekele2023}, HCI scholars should address existing inequalities in software with global popularity. We encourage researchers to engage with communities prior to embarking on this work, rather than assuming that Indigenous people are interchangeable.\footnote{Community concerns about the imperial language, Russian, guided our audit of YouTube in Kyrgyzstan, so we attended to linguistic hierarchies in our analysis. However, repeating our algorithmic audit in Algeria, we found that YouTube most frequently recommended Quranic \textit{surah} recitations, not French-language content, so we could not meaningfully interpret these results absent additional knowledge about local concerns.} Power relations manifest differently across contexts; more work is needed to elaborate the specific harms faced by Indigenous populations. We must consider how algorithmic curation serves to amplify imperial influence, reimposing oppressive power structures that decolonial efforts have sought to dismantle.

Our work makes three main contributions:
\begin{itemize}
\item Demonstrating how an online platform can propagate colonial cultural dominance, rather than revitalizing Indigenous cultural heritage.
\item Connecting the online amplification of colonial content to offline uptake of colonial ideologies.
\item Equipping communities with end-user strategies to reduce imperial amplification and align recommendations with Indigenous values.
\end{itemize}

\section{Related Work}
\label{sec:related}

Our work applies theoretical decolonial frameworks to technical algorithmic auditing methods. We first describe decolonial computing, which centers the priorities and values of Indigenous populations to reorient lines of inquiry. Then, we sketch the existing landscape of algorithmic auditing, drawing attention to the overemphasis on Western concerns in Western contexts. In both areas, we provide examples of research that has pursued the interests of Indigenous populations, though no algorithmic audit has yet assessed how algorithms disseminate Indigenous content.

\subsection{Decolonial computing}

Colonialism is a historical expansionist project in which one political power dominates another land and its peoples through settlement, military strength, and economic control to expropriate resources \cite{ali2016}. This process produces difference between the colonizers and the Indigenous populations of the colony, where local cultures, relations, and people are cast as inferior \cite{tlostanova2012, fanon2007}. Even when an imperial power ends its formal administration of a colony, colonial power relations continue to shape the culture and consciousness of formerly colonized peoples \cite{tlostanova2012, Abashin2023}. For instance, many people in former colonies still apply racial classification schemes that originated in the imperial core (e.g., sorting people as ``white'' or ``Black''), which demonstrates that discriminatory colonial ideologies continue to shape how formerly colonized people understand and act in the world \cite{tlostanova2012}. Intellectuals and activists have articulated and advanced decolonial theory to confront these lingering effects of colonialism.

Decolonial thinking aims to reduce the influence of colonial hierarchies and stereotypes. Colonialism presents a totalizing worldview, creating a scale where colonizers are located closer to full humanity and the colonized nearer to barbarism, justifying colonial incursions as civilizing missions that advance medicine, standards of living, and development \cite{fanon2007, cesaire2001}. A decolonial response might challenge how colonizers apply the term ``barbaric'' to others, given the plundering, torturing, and killing that they perpetrate in the colonies \cite{cesaire2001}. Rather than relying on concepts developed in the imperial core, the decolonial approach emphasizes knowledge developed to serve people in the periphery. This effort does not require rejecting all existing forms of knowledge, but calls to reorient its use to serve Indigenous views and needs \cite{tlostanova2012, torres2017}.


As technical systems designed in the core are often deployed in the periphery, scholars have called for computing researchers to adopt decolonial thinking when evaluating impacts \cite{garcia2021, ali2016, lazem2022}. HCI scholarship often develops universal methods and standards by working with populations from the core; such methods cannot account for local context and may not generalize to Indigenous populations \cite{garcia2021, lazem2022, dourish2012, irani2010}. Funding structures dictate who gets to perform research \cite{nocera2024}, and international partnerships often frame researchers from rich countries as knowledge producers and Indigenous partners as data collectors or translators \cite{talhouk2023, lazem2022}. These dynamics prompt concerns about extractivist research, where knowledge gathered from a local community is transformed into a research contribution, without investigating participant concerns or communicating back results \cite{garcia2021, millan2024, lazem2022}. 

Decolonial computing centers knowledge generated by and for the periphery to advance the self-determination of Indigenous people, embracing many cultures and perspectives \cite{garcia2021, nocera2024, lazem2022, ali2016, millan2024}. This orientation does not require complete rejection of existing technologies; Indigenous and Western knowledge can both inform decolonial computing efforts \cite{lazem2022, millan2024}. There are many paths forward: some scholars promote reflexivity to combat harms \cite{talhouk2021}, others propose selecting methods based on their potential benefit to the population that the research serves \cite{garcia2021}. Decolonial computing projects can take many forms, from interventions in HCI curricula \cite{peters2016}, preservation of cultural heritage \cite{lazem2022, millan2024}, futuring through Indigenous frames \cite{millan2024}, and appropriating tech for Indigenous purposes \cite{kotut2022, millan2024}. In addition to envisioning future technology designs and practices, there have been efforts to identify harms from colonial logics embedded in algorithms and to propose improvements to enhance sovereignty. 

Past research has documented several ways that algorithms can reinforce colonial logics. Some work identifies explicitly extractive historical systems used by colonial administrators, which reallocated resources from the periphery to the core \cite{Dryer_2023, Thorat_2019}. HCI research on modern digital platforms often focuses more on the creation of scales of social difference rather than definite material redistribution, with emphasis on the portrayal of language, culture, race, ethnicity, religion, and gender \cite{Das_2024, Almeida_2024, Shahid_2023, Pendse_2022, Nigatu_2024, mim2020, chen2023, das2021, dosono2020}. For example, Bangladeshi Facebook users perceived content moderation practices on the platform as reinscribing colonial relations that privileged Western norms and unfairly censored minoritized views \cite{Shahid_2023}. Some suggested solutions to counteract colonial logics within algorithmic systems involve diversifying development teams \cite{Das_2024}, improving regulation of platforms \cite{Nigatu_2024}, and fully redesigning existing systems \cite{Pendse_2022, Shahid_2023}. Much of this work acknowledges the difficulty of changing the practices of global capitalist firms; thus, we build upon this broad project of decolonization in HCI by contributing end-user strategies for decentering imperial culture that do not require platform cooperation.

\subsection{Scope of algorithmic auditing research}

When algorithmic systems structure how users access and make sense of information, algorithmic biases can distort how people understand the world \cite{noble2018}. These distortions can be harmful; one study found that Google promoted arrest record websites in search results for African American-sounding names but not white American-sounding names, reinforcing negative stereotypes that link African Americans with criminality \cite{sweeney2013}. Users often internalize personalized algorithmic recommendations \cite{colbjornsen2018}, even though they have little power to contest the inferences made by recommender systems \cite{gaw2022}. Given the potential of recommender systems to influence user beliefs, algorithmic auditing is a vital method for probing prioritization and curation of content for a range of possible biases \cite{Trielli2019, sandvig2014auditing}. 

Algorithmic audits evaluate black-box systems like YouTube, whose internal workings are not directly inspectable by external researchers. The most common algorithmic auditing approach varies the inputs to a system and records its outputs \cite{metaxa2021}. For instance, researchers tested recruiting software for discrimination by submitting similar resumes but varying the names of applicants, finding that names associated with Black men are greatly disfavored \cite{wilson2024gender}. Though this approach does not reconstruct internal processes, it can be used to detect problematic patterns in algorithmic behavior. 

Yet, some forms of bias have received much more attention than others, as the study design of an algorithmic audit often specifies a limited set of biases for examination. For instance, by simulating users with different characteristics (e.g., varying race, gender, or family status), researchers can record and analyze the outputs of algorithmic systems to identify disparities in the content recommended to different user groups \cite{Asplund2020, escher2020}. In such a setup, researchers must specify the biases under examination prior to executing their studies, rather than allowing patterns of harm to reveal unanticipated types of bias. Past audits have tested whether particular recommender systems amplify politically radical content  \cite{ribeiro2020, Ledwich_Zaitsev_Laukemper_2022, haroon2022youtube, Makhortykh_Urman_2020, haroon2023auditing, heuer2021, lutz2021}, gender prejudice \cite{Eriksson_Johansson_2017, vanes_2021, urman2024foreign, hannak2017, metaxa2021image, kay2015}, anti-immigrant sentiment \cite{urman2022auditing, Ieracitano2024}, racist stereotypes \cite{sweeney2013, noble2018, metaxa2021image, hannak2017}, misinformation \cite{tomlein2022, srba2023, abulfottouh2020}, conspiracy theories \cite{hussein2020, faddoul2020, schellingerhout2023}, and religious intolerance \cite{albadi2022, moe2019}. However, markers like race are most frequently examined through U.S. classification schemes (e.g., ``White vs. non-White'' \cite{urman2024}), though different racial categories are salient in other cultural contexts \cite{abdu2023}. Scholars have cautioned that Western paradigms can break down in non-Western contexts \cite{ramesh2022}, so formulations of bias that employ Western categories may not extend to non-Western contexts \cite{metaxa2021, urman2024}.

As Western academics generally set the agenda for the issues that algorithmic audits pursue, concerns from outside the Western frame typically go uninvestigated. The vast majority of algorithmic auditing studies took place in the US or Western Europe \cite{urman2024} and emphasize Western consumption patterns that do not hold globally \cite{metaxa2021}. Most studies that include data from countries outside of Western liberal contexts are ``larger-scale comparative analyses'' that do not focus on problems specific to the studied contexts \cite{urman2024}.\footnote{This Western skew is not unique to algorithmic audits, but marks a general trend in computing literature. Scholars have argued that study findings greatly overrepresent people who are WEIRD---Western, Educated, Industrialized, Rich, and Democratic \cite{henrich2010a} (e.g., 73\% of CHI papers include exclusively participants from Western contexts \cite{linxen2021}, and 84\% at FAccT \cite{septiandri2023}). Yet, only 12\% of the global population lives in WEIRD societies, and results cannot be presumed to generalize to other varied populations living in diverse contexts \cite{henrich2010}.} As companies have addressed biases identified by algorithmic audits \cite{raji2019}, it is likely that any benefits from technical improvements are concentrated in Western contexts.

A handful of recent algorithmic audits have challenged or transcended Western frames. Researchers have charged recommendation platforms with prioritizing content produced in America \cite{gomez2021} or in the English language \cite{rovira2021}. Further, the mechanisms that Western recommender platforms use to protect users from harm can break down in non-Western contexts, with weaker enforcement of policies for low-resource languages \cite{nigatu2024} and in countries outside the global core \cite{silva2020}. Some algorithmic audits have examined how imperial powers can promote particular narratives through dissemination of content and manipulation of search rankings \cite{toepfl2023}; for instance, Yandex, a Russian-based search engine, suppresses results that label Soviet-era famines as genocides \cite{makhortykh2022}. American platforms can also reinforce imperial narratives; one study found that Google Maps redraws country borders based on geolocation: users in Russia see the contested Russian-occupied territory of Crimea with a line severing it from Ukraine, but this line is not present on maps accessed from other countries \cite{soeller2016}. Building upon this critical work, our study explicitly applies a decolonial perspective to algorithmic auditing, examining whether platform recommendations prioritize imperial content in former colonies. 

\section{Positionality}

The authors come from several different national origins---American, Serbian, and Kyrgyzstani---but we are all associated with institutions that have been shaped by imperial ideals of education. The second author is affiliated with a school in Bishkek that receives significant funding from the United States and whose primary language of instruction is English, and the other authors are affiliated with a North American research university. Though our research team primarily coordinated work in English, among the authors, two speak Kyrgyz, and three speak Russian. We recognize that our position as researchers gives us greater power to speak than community members and that we bring in our own interests when interpreting and spotlighting their contributions. While these power imbalances are inextricable from our study, we aim to make this research useful to caregivers in the community, motivated by an ethos of care and solidarity.

\section{Kyrgyz Case Study Background}
\label{sec:background}

Though rarely listed alongside Britain, Spain, and France as a major colonial power, Central Asian scholars have argued that Russia ought to be understood as an empire \cite{Kassymbekova_2023, Kassymbekova_Marat}. They frame recent violent expansions into former Soviet republics, such as Georgia and Ukraine, as modern manifestations of a centuries-old trajectory of imperialism \cite{Kassymbekova_2023}. Ambitious officers in the tsarist Russian Empire (1721--1917), eager to prove Russia as a European power, initiated a series of incursions into Central Asia \cite{Morrison_2020}. After bloody conquest and violent suppression of rebellions, Russian politicians considered the region ``as a colony analogous to French North Africa or British India [...] subject to elaborate plans for economic exploitation and increased Russian settlement’’ \cite{Morrison_2020}. The sense that this predominantly Muslim area was ``backward'' and required civilizing through European norms persisted into the Soviet era \cite{northrop2004veiled, Sahadeo_2016, Kassymbekova_Marat}.

The colonial nature of the USSR (1917--1991) is debated; however, like any regime termed ``colonial,'' policy is ``complex and hybrid,'' never conforming to idealized models \cite{Abashin2023}.  Examining the ongoing colonial/neocolonial influence of the Soviet Union and Russia is an analytic tool to understand the effects of forced assimilation, economic exploitation, and political subjugation perpetrated in the region and how Central Asian people conceptualize their relationship to the Soviet past \cite{Abashin2023,  Chari_Verdery_2009}. While our analysis terms the USSR colonial, we acknowledge and explore how decolonialism, and Soviet memory, in Kyrgyzstan is fraught. 

The early Soviet policy of \textit{korenizatsiia} seemingly promoted and valued difference, but later policies promoted a unified Soviet identity tied to Russian language and culture \cite{martin2001}. Though titular nationals were preferentially hired within their republics, advancement required knowledge of Russian and cultural assimilation \cite{2000_Hirsch, martin2001, Slezkine_1994, smith2006non}. The Soviet Union continued Central Asia's economic exploitation and ecological devastation started by the Russian Empire---compelling mobile-pastoral groups to adopt Slavic agricultural practices and ignoring local needs even during famines \cite{Loring_2014, graziosi2022genocide}. In the republics, Russian became mandatory as a second language in school, and academia throughout the USSR functioned primarily in Russian \cite{Kirkwood1990LanguagePI, grenoble2003language, fierman1991language}. As a result, Russian remains a majority language in many former Soviet republics. Though Kyrgyzstan declared its independence from Moscow in 1991, it has elevated Russian to an official language \cite{mcdermott2019landscape}.

The USSR's modernizing agenda and pro-nationalities policy shapes memory of the Soviet Era and feelings toward the modern Russian Federation. Nested relationships of hierarchy led to different experiences of domination from both the core and peripheries, complicating the narrative of Soviet colonial subjugation and how non-Russian groups in post-Soviet Central Asia conceptualize their identity as Indigenous, national, or somewhere in-between \cite{goff2021nested}. In Central Asia, Russia is held in high esteem, with better favorability ratings than the U.S. and China \cite{laruelle2020no}. Describing Russian rule of Kyrgyzstan as ``colonial'' is controversial even within supposedly liberal academic spaces, with students working on topics related to colonialism in the region reporting push-back from administrators \cite{alybaeva2023}. 

The pressure to avoid discussions of Soviet colonialism is due to disparate understandings of the Soviet past as modernizing or colonial, and the power the Russian Federation still holds over the region. Russia exercises soft power in Central Asia through the market, with widespread popularity of Russian pop culture products, and official government policies, such as interventions in school curricula to promote Russian language and narratives or suspension of imports in reprisal for language policy seen to be ``anti-Russian'' \cite{Nourzhanov_Peyrouse_2021, marat2023reclaiming}. Russia also wields economic power through migrant travel to Russia in pursuit of higher salaries, where workers face open discrimination and harassment \cite{Nourzhanov_Peyrouse_2021, Sahadeo_2016}. Despite Russian influence and conflicting Soviet memory, Kyrgyz decolonial activism and decolonial discourse has become ``mainstreamed'' after the Russian invasion of Ukraine and engagement with post-Soviet decolonial content on platforms such as Instagram and TikTok \cite{alybaeva2023, marat2023reclaiming}.

In Bishkek, the decision to use Kyrgyz or Russian is laden with moral and political considerations, and further complicated by anxieties over presumed Kyrgyz loss among younger speakers. While adults largely believe that Russian is essential for upward social mobility and economic success, children who speak primarily Russian are derided as ``little Russians'' and thought to exhibit qualities attributed to ethnic Russians \cite{korth2005language, pavlenko2006russian, mcdermott2019}. Despite social stigmatization of the purportedly ``Russianized'' children of urbanites, monolingual Kyrgyz speakers are stereotyped as backward, rural dwellers, and uncultured. Children in some middle-class households are growing up to be primarily Russian speaking, which is tacitly reinforced by their parents speaking to them largely in Russian \cite{mcdermott2019}. This pattern suggests a disruption of intergenerational transmission and thus indicates a shift in progress toward Russian in the city \cite{eberhard2021ethnologue, fishman1991reversing,
Grenoble2005saving, unesco2003}. 

Bishkek residents report anxiety over the status of Kyrgyz, but do not explicitly name the fear ``language endangerment'' or ``shift''---likely as the terms imply colonial legacies \cite{meek2019language}. The language issue has been taken up by both government entities and language activists, who promote Kyrgyz through language maintenance efforts: translating Wikipedia articles; adding Kyrgyz to Google Translate; requiring Kyrgyz on official documents; mandating language tests for government officials;  publishing Kyrgyz textbooks; and sharing language-learning content on social media \cite{derbisheva2009, huskey1995politics, korth2005language, alybaeva2023}. Government-led initiatives, however, are generally unsuccessful at maintaining Kyrgyz due to incomplete implementation, lack of financial support for language programs and the assessment of programs, and changes in programs due to fear of political reprisal, either internally or from Russia. While many language maintenance efforts do little to encourage more children to learn Kyrgyz, there is a growing interest among Kyrgyz adults in maintaining Kyrgyz through family language policies and adult language learning.

\section{Prior Case Study Findings}
\label{sec:casestudy}

In this paper, we conduct an algorithmic audit informed by our prior ethnographic work \cite{mcdermott2026}; this section excerpts the findings that motivate our current inquiry. The goal of our previous research was to 1) investigate language change in Bishkek, and 2) explore the social factors driving change that may not be accounted for by traditional measures of language endangerment. While this work uncovered serious caregiver concerns about the impact of children's technology practice on their language preferences, the ethnographic methodology could not empirically confirm their suspicions about the YouTube platform or arm them with practices for contestation. 

The ethnographic methods in our prior work consisted of participant observation, semi-structured interviews, and recordings of quotidian speech within Bishkek families. We spent approximately eight months working with 15 different ethnic Kyrgyz families, who were recruited through network sampling ~\cite{heckathorn2017network}. The families included 46 children from the age of 9 months to 16 years. These families, who had relocated to Bishkek within two generations, were observed in various daily activities, including meals, homework, and media interactions. The data collected, including over 100 hours of audio and video recordings, was analyzed using conversation analysis, discourse analysis, and thematic analysis.

The following subsections highlight the specific findings that guide our algorithmic auditing inquiry. Caregivers are concerned about children's preference for Russian, perceiving this shift as a threat to Kyrgyz identity. They often blame online platforms, most notably YouTube, for shaping their children's preferences (YouTube is the second most visited website in the country after Google \cite{alexa}). However, they are unsure how to contest the power of platforms and have yet to identify successful strategies that increase their children's interaction with Kyrgyz-language content. Our algorithmic audit thus seeks to confirm that the YouTube platform steers children toward non-Indigenous content and test strategies for aligning recommendations with caregiver preferences. By integrating prior ethnographic insights with algorithmic analysis, we aim to provide empirical evidence and practical solutions for preserving Kyrgyz language and culture.

\subsection{Children's Language Practices Differ from Adults}

Broadly, our prior study found that language shift in Bishkek was more pronounced for children than adults. Adults mix Kyrgyz and Russian frequently, while children do not. Many children exclusively spoke Russian, even in response to Kyrgyz utterances. Though some young adults are re-learning Kyrgyz, either because they see it as integral to their identity as Kyrgyz people or because of decolonial intentions (or both), children disprefer Kyrgyz to Russian, and have little interest in Kyrgyz study. Children in interviews explicitly acknowledged feeling ashamed of the way they speak Kyrgyz, and discomfort when attempting to speak the language. 

\subsection{Caregivers are Concerned that Language Shift Threatens Kyrgyz Identity}

Within Bishkek, adults fear that loss of the Kyrgyz language means a loss of Kyrgyz identity; loss of Kyrgyz traditions, history, and ways of being \cite{mcdermott2019, meek2022}. In interviews, some parents stated they felt their children spoke ``like foreigners.'' Adults chastised children for not knowing Kyrgyz well enough with the remark ``Are you even Kyrgyz?'' which draws upon discourse that language is tied to identity. As one participant explained during participant observation:
\begin{quote}
She said, ``if you lose your language, you lose your identity.'' She told again the story of her neighbor whose kids all moved to the United States. None of them knew the proper traditions to carry out after his death. They didn't read Quran at his grave, and they couldn't appropriately divide the horse at the funeral dinner. ``In their souls they weren't Kyrgyz---this is why it is important to know your own language. The family couldn't show respect. Language is such a power.'' Fieldnotes 3/20/23
\end{quote}

\subsection{Caregivers Attribute Language Shift to Influence of YouTube}

Over the course of participant observation and interviews, caregivers and educators explicitly named YouTube as one of the leading causes of children's preference for Russian. We found strong supporting evidence for their beliefs. Children spent a significant amount of their time engaging with Russian-language content on YouTube and incorporated that content into their offline behaviors.

During participant observation, children spent more time watching videos on YouTube than using TikTok, playing phone games, browsing social media, or creating their own content. In interviews parents and caregivers reported that their children spent at least 2 to 3 hours per day engaged with online media; however, due to stigma surrounding screen time, it is likely this number is much higher (see \cite{livingstone2020parenting}). Observations of children watching YouTube indicated that a high proportion of content consumed was in Russian: 70 of 74 videos, with one video in English and three with only instrumental music. Since YouTube viewing was frequently a group activity; children were socialized into prolonged YouTube use from a very young age. Kids collaboratively entered search terms, and even preverbal toddlers helped select videos by pointing at thumbnails.

Children’s engagement with YouTube extended into their offline lives, as they frequently discussed videos and mimicked language heard online. Children notably used features of Russian speech that they are exposed to through YouTube---most salient of these were lexical items and phrases associated with their favorite bloggers and videos. For example, the Russian-language fan-made video ``\foreignlanguage{russian}{ДЕВОЧКА УЭНСДЕЙ}/Girl Wednesday'' was particularly popular, with children singing and quoting it frequently during participant observation.

Even in the regions, where Kyrgyz is the majority language, people report knowing children who began speaking Russian after exposure to Russian-language content via YouTube. During participant observation, many even gave anecdotes of children who use English to demonstrate the culmination of what they see to be the pernicious influence of YouTube on children's linguistic practices. For example, one government employee described in two different interviews how her granddaughter, who lived in a village in Chui Oblast, could name all of her colors in English and Russian, yet not in Kyrgyz. She noted that the girl was always on her phone, and suggested that if her mother did not have to work such long hours she would likely speak more Kyrgyz. 

\subsection{Caregivers Unsure How to Contest Influence of Platforms}

Some caregivers attempted to modify the behavior of online platforms, but their strategies were rarely successful because they did not know what technology practices would achieve their desired states. For instance, one family blocked popular Russian bloggers in the hopes YouTube would suggest more Kyrgyz content. Instead, they reported that their children now watched content almost exclusively in English. While these efforts changed the videos recommended to children, they did not boost the proportion of Kyrgyz recommendations in line with caregiver desires. 

Another approach was to restrict access to devices, but these strategies were typically impractical due to high demands on caregivers. When not in the hands of children, adults hid phones, remotes, tablets, and laptops on top of refrigerators, bookshelves, and wardrobes. Other caregivers tried to impose screen time limitations, such as only 30 minutes a day after finishing homework. However, caregivers struggled to apply these practices consistently. In many households, one or both parents are absent as migrant workers, with as much as 20\% of the Kyrgyz population working abroad as of 2022 \cite{iom2022}. Other familial duties also make it difficult for parents, especially young mothers, to supervise their children. Wives are often expected to care for their in-laws and their own parents, all while maintaining what are seen to be traditional practices of hospitality that include accommodating guests with elaborately laden tables and tea service. Though adults in Kyrgyzstan are immersed in the global discourse that screen-time is bad for their children, Bishkek kids still spend much of their time on devices. 

\section{Algorithmic Auditing Method}
\label{audmeth}

Our prior ethnographic inquiry uncovered concerns about YouTube's influence, but caregivers wanted to know more about how the platform operated. Did YouTube steer their children toward non-Indigenous content? Given their interest in this question, our algorithmic auditing inquiry investigates the language of platform recommendations to inform caregiver beliefs and responses to YouTube. Importantly, if the platform was funneling their children toward Russian-language videos, caregivers wanted to know if anything could be done to counteract this amplification of colonial content. Given that broad efforts to restrict children's use of devices have failed, our algorithmic auditing inquiry seeks to identify targeted practices that can impact the language of the content presented by platforms.

To address these caregiver concerns, we framed a series of research questions to describe the proportion of languages (Kyrgyz, Russian, English, and others) in the content that YouTube promotes under different conditions and for different user groups. Each of our three research questions corresponds to different ways participant children selected videos: choosing content from recommended videos on the platform homepage, searching for topics related to their interests, and tapping on suggestions displayed while browsing. We also investigate strategies Kyrgyz users speculated could align the recommendation algorithm with Indigenous intentions. These research questions evolved through conversations between the American and Kyrgyz authors, Kyrgyz research assistants, and participant families.

\begin{enumerate} [ {RQ}1:]
	\item What content does YouTube present to new users across Kyrgyzstan?
	\item What content does the YouTube search algorithm prioritize in Kyrgyzstan?
	\item How do the recommendations encountered by Kyrgyz children browsing the platform differ from those promoted to Kyrgyz adults?
\end{enumerate}

We elected to investigate these questions with a sock-puppet audit, a method effective in uncovering bias in YouTube recommendations without collecting sensitive user data. Most algorithmic audits require running numerous tests to detect bias; to scale such efforts, the sock-puppet approach uses computer programs to simulate user activity \cite{sandvig2014auditing}. Sock-puppets are programmed to perform automated actions like opening a webpage, scrolling, and clicking on links. Rather than conducting ethically-thorny wide-scale surveillance of Kyrgyz children using YouTube, our sock-puppets emulate users with different features, perform activity on the platform, and record resulting platform recommendations. We also selected the sock-puppet auditing method because it can determine whether signaling an interest in certain types of content will result in a ``filter bubble,'' where the algorithm promotes many similar videos that can serve to reinforce user biases and ideologies \cite{hussein2020, srba2023, haroon2022youtube, albadi2022}. We were interested to learn when certain activity on the platform led users to experience a Russian-language filter bubble. Finally, we selected this method because prior YouTube audit designs had also investigated the three user behaviors featured in our research questions: opening the homepage, searching for terms, and browsing the platform \cite{hussein2020, juneja2023, schellingerhout2023, albadi2022, spinelli2020, Papadamou2022, tomlein2022}. We investigate all three pathways, as prior work has indicated that different YouTube recommendation sources can highlight different sets of content \cite{hussein2020}.

In this section, we outline the procedure for creating sock-puppets and collecting video recommendations, then apply these techniques in a series of experiments that address our research questions. Finally, we describe how we annotated the languages in our video observations.

\subsection{Creating Sock-Puppets and Collecting Video Recommendations}
\label{collecting_vids}

We use sock-puppets to interact with the YouTube web user interface to collect all video recommendations. Our sock puppets are programmed to emulate user behaviors---such as opening the YouTube homepage, selecting video recommendations, searching for terms, and watching videos---and to scrape and record all video recommendations encountered.  Each sock-puppet is implemented through an automated Firefox web browser instance that sets Kyrgyz as a language preference and accesses YouTube in an isolated container. We implement our sock-puppets in Python, use the Selenium Python bindings and Selenium Firefox WebDriver to interact with the YouTube website and scrape recommended video information, and insert scraped data into a SQLite database. To prevent carry-over effects between executions \cite{hannak2013}, each sock-puppet is only used once. We run our experiments on machines physically present in Kyrgyzstan and connected to public network access points.

\begin{figure}[h!]
\begin{minipage}[b]{0.45\linewidth}
    \begin{subfigure}{\linewidth}
\includegraphics[width=\textwidth]{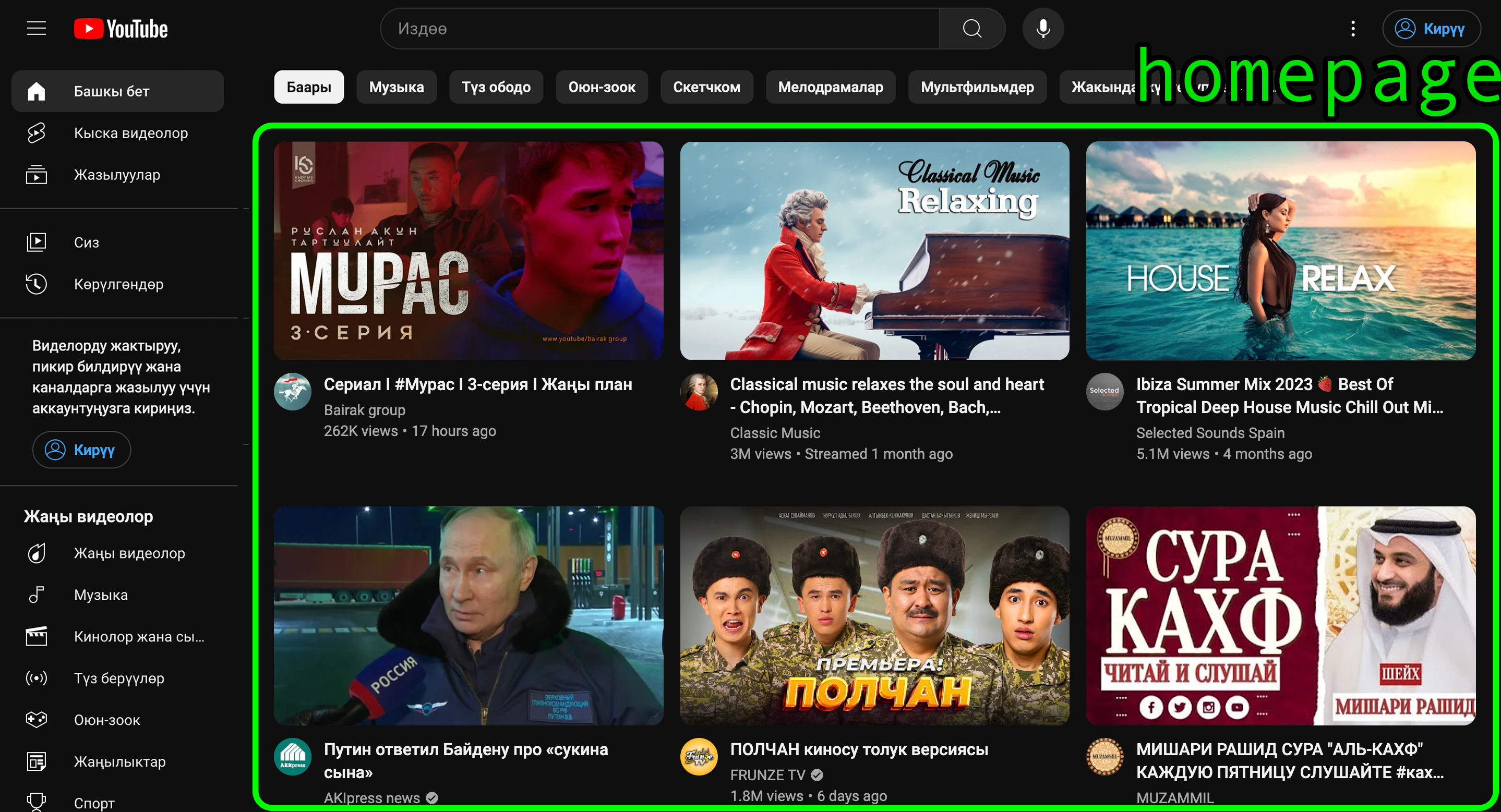}
        \caption{}
        \label{fig:a}
    \end{subfigure}

    \begin{subfigure}{\linewidth}
         \centering
         \includegraphics[width=\textwidth]{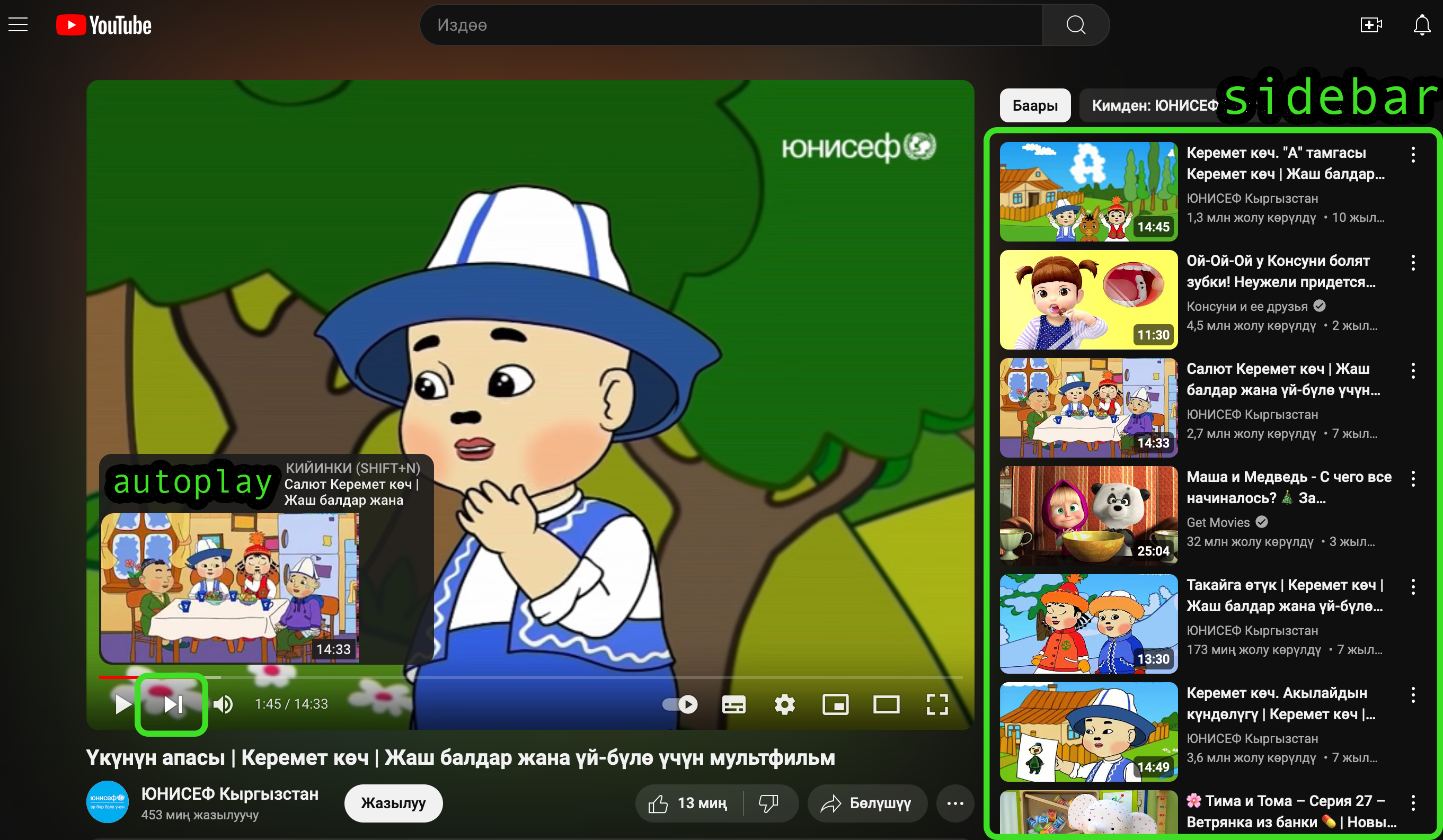}
         \caption{}
         \label{fig:interference}
    \end{subfigure}
\end{minipage}%
\hfill
    \begin{subfigure}{0.5\linewidth}
        \centering
        \includegraphics[width=\textwidth]{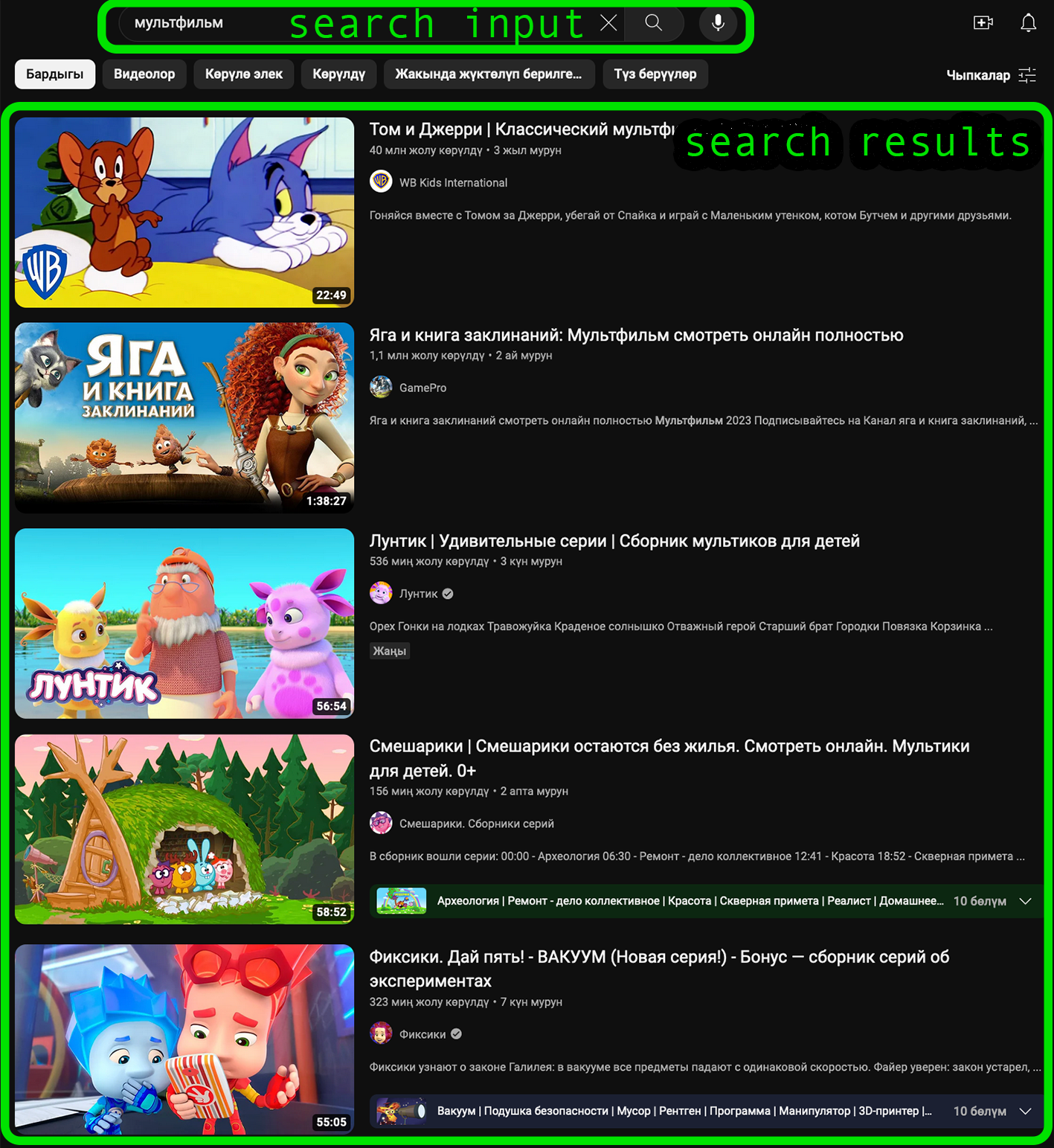}
        \caption{}
        \label{fig:Bragg}
    \end{subfigure}
\caption{Screenshot of the YouTube homepage (a), video page (b), search page (c).}
  \Description{Screenshots of the YouTube homepage, search result page, and video page.}
	\label{fig:rec_sources}
\end{figure}

We collect data from four different recommendation sources across YouTube: 

\begin{itemize}
	\item Homepage recommendations, displayed in a grid on the platform homepage 
\item Search results, returned in response to a query
\item Autoplay videos, automatically loaded after a video is watched
\item Sidebar suggestions, listed to the right of an active video (refer to Figure \ref{fig:rec_sources} for the location of each recommendation source on the platform). 
\end{itemize}

\subsection{Experiment 1: What content does YouTube present to new users across Kyrgyzstan?}
\label{RQ1}

\begin{figure}[h!]
  \centering
  \includegraphics[width=0.75\linewidth]{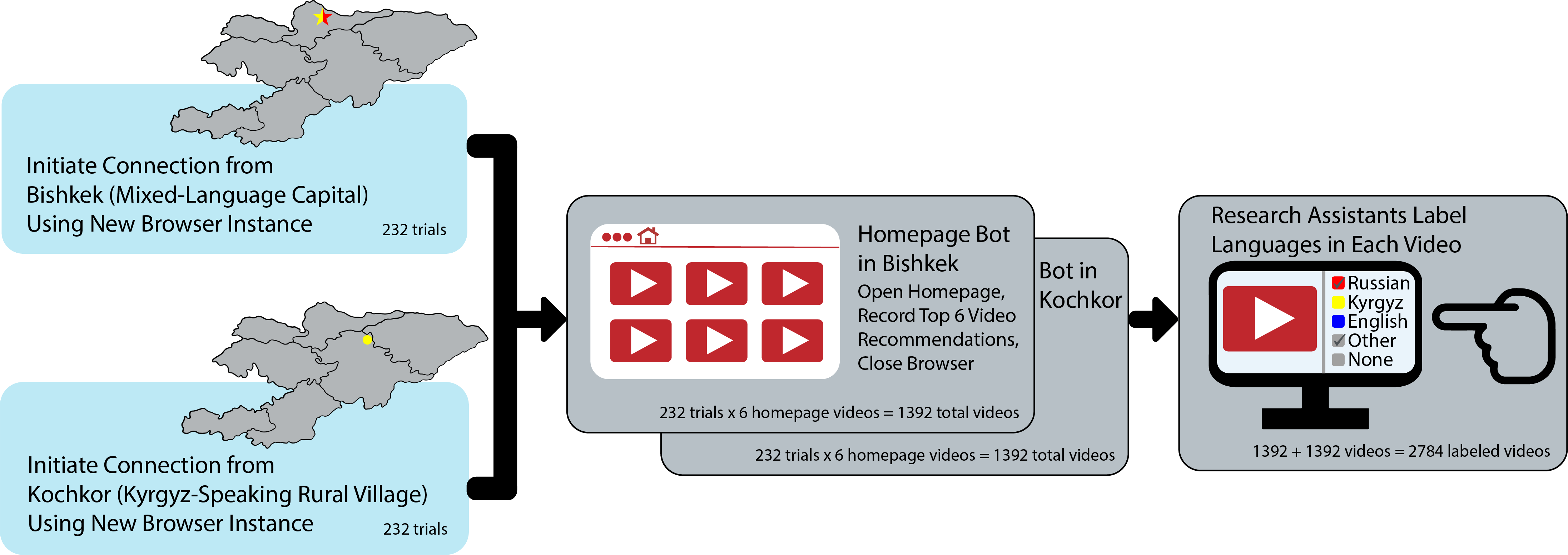}
  \caption{Steps performed in Experiment 1 by the homepage bots. Two locations are evaluated: the capital city of Bishkek and the village of Kochkor.}
  \Description{}
	\label{fig:rq1_experiment}
\end{figure}

Our first experiment aimed to describe the video recommendations presented to new users on the YouTube homepage in Kyrgyzstan. Each sock-puppet opened the YouTube homepage and recorded the top six video recommendations. This experiment reflects what a person would see when they first access the platform, prior to watching videos and signaling their interests to the algorithm. 

Building on ethnographic accounts of children in rural areas learning Russian via YouTube, we tested whether geographic location influences the language of video recommendations. We concurrently executed 232 sock-puppet trials from a public network access point in Bishkek (the country capital) and 232 trials in Kochkor (a rural village). Though the Russian language has had a longstanding historical presence in Bishkek, other areas of the country have much lower exposure to Russian. Kochkor is a village located 200 km southeast of Bishkek in the Naryn region. Russian is rarely spoken in Kochkor, as the population in the region is 99.2\% ethnically Kyrgyz and reports only 0.06\% native Russian speakers \cite{kochpop2010}. In contrast, Bishkek's population is 66.2\% ethnically Kyrgyz and 23.2\% identify as native Russian speakers \cite{bishpop2010}.

If the YouTube algorithm reflected local language customs, we would expect video recommendations in Kochkor to be predominantly Kyrgyz, while Bishkek would feature recommendations in a mix of languages. However, given the findings of our ethnography, we hypothesize that the YouTube algorithm violates local language customs, disproportionately recommending Russian-language content.

\subsection{Experiment 2: What content does the YouTube search algorithm prioritize in Kyrgyzstan?}

\begin{figure}[h!]
  \centering
  \includegraphics[width=\linewidth]{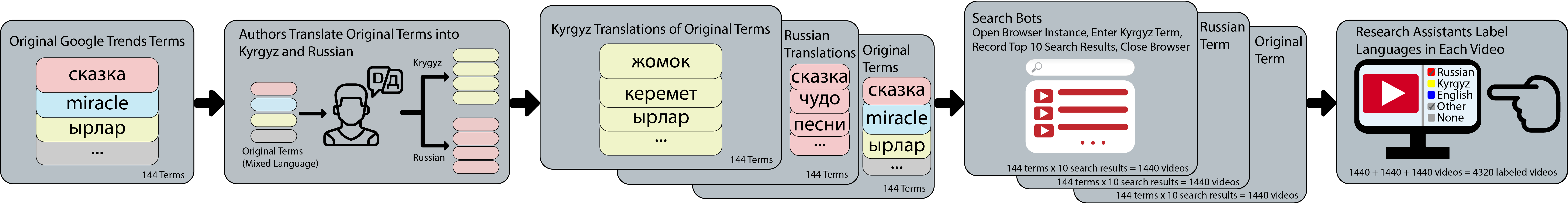}
  \caption{Steps performed in Experiment 2 by the search bots. Three sets of terms are evaluated: original search terms from Google Trends, Kyrgyz translations, and Russian translations.}
  \Description{}
	\label{fig:rq2_experiment}
\end{figure}

Our next experiment aimed to determine how the YouTube search algorithm responded to different forms of inputs. We first describe our steps for constructing our set of search inputs: compiling popular search terms from Google Trends, checking them against our ethnographic observations, and translating them into different languages. Finally, we describe how our sock-puppets used these terms to collect video recommendations from the search algorithm.

\subsubsection{Compiling Search Terms from Google Trends}

To identify popular YouTube search terms in Kyrgyzstan, we use Google Trends, which is a tool that presents interest in search queries over time and across regions, interest categories, and platforms. In May 2023, we extracted the top and trending search terms in Kyrgyzstan on the YouTube Platform for several time intervals (5 years, 1 year, 90 days, and 30 days). We gathered this data for both the ``All categories'' and ``Children's Interests'' categories. From these data sources, we compiled the terms and sorted them into three lists: mixed interest (terms appearing in both the categories), adult-interest (terms appearing only in ``All categories''), and child-interest (terms appearing only in ``Children's Interests''). This resulted in 17 mixed-interest search terms (e.g., ``\foreignlanguage{russian}{музыка}''/music\footnote{The term in quotes is an observed search, and an English translation follows after the slash.}), 85 adult-interest search terms (e.g., ``\foreignlanguage{russian}{машина}''/car), and 59 child-interest search terms (e.g., ``\foreignlanguage{russian}{сказка}''/fairy tale), for a total of 161 unique search terms. We list the child-interest terms in Appendix \ref{appendix:childterms} and adult-interest terms in Appendix \ref{appendix:adultterms} (the mixed-interest terms are not tested in this study; we compiled and excluded them to reduce overlap of the child and adult categories).

\subsubsection{Validating Google Trends Data with Ethnographic Observations} 

Though Google Trends produced a representative set of search terms for our target region \cite{jun2018}, we sought to validate the interest categories using our ethnographic observations. As Google does not provide the basis for their bucketing of search terms into different categories and offers uneven support for the region\footnote{E.g., many Google APIs with fields for geographic location did not include Kyrgyzstan as an available region.}, we had reservations about accepting Google's privately constructed categorizations. Thus, we aimed to confirm that the ``Children's Interests'' category matched what we knew about the consumption behavior of Kyrgyz children. Using direct observation and device histories, we formed a list of search queries from the ethnographic data. When we collected device history, we asked adult caretakers whether each term was related to their interests or their children's. Of 59 ``Children's Interests'' searches from Google Trends, we could corroborate 35 from ethnographic observations. The remaining Google Trends searches, though not observed, seemed plausible (e.g., ``\foreignlanguage{russian}{винни пух}''/Winnie Pooh and ``\foreignlanguage{russian}{конек горбунок}''/The Little Humpbacked Horse did not appear, but they are instances of observed searches for genres ``\foreignlanguage{russian}{мультики}''/cartoons and ``\foreignlanguage{russian}{сказки}''/fairytales). Our institutional review board approved our research design (HUM00197234). 

Though the ethnographic searches could serve as a basis for a search query set, we opted to use the Google Trends data as our ethnography participants were recruited through networks, which may not have yielded a representative sample~ \cite{heckathorn2017network}. By validating Google Trends data with ethnographic observations, we ensure that our search queries reflect the actual interests of Kyrgyz children while maintaining a representative and broadly applicable dataset.

\subsubsection{Translating Search Terms into Different Languages}

To compare how the search algorithm treated different input languages, we translated the terms into Kyrgyz, Russian, and English (English translations are included for reader comprehension but not tested). An author fluent in these three languages wrote the equivalents of each term, and another author with language proficiency confirmed the translations. We also noted that some terms had the same translations into Kyrgyz and Russian, which we mark as bivalent in Appendices \ref{appendix:childterms}
 and \ref{appendix:adultterms}. We further observed that the terms were often typed informally, substituting Kyrgyz letters with Russian Cyrillic counterparts that are more accessible on standard keyboards. We report on the impact of this typing practice in Appendix \ref{appendix:typing}.

\subsubsection{Search Algorithm Experimental Setup}

We investigated the original 144 search terms from Google Trends (59 child, 85 adult) and their translations into Russian and Kyrgyz (totaling 432 searches). Each sock-puppet opened the YouTube homepage, submitted a single term to the search bar, then recorded the top ten video search results. We performed these trials in Bishkek and parallelized formulations of searches for the same concept to reduce temporal effects. 

\subsection{Experiment 3: How do the recommendations encountered by Kyrgyz children browsing the platform differ from those promoted to Kyrgyz adults?}

\begin{figure}[h!]
  \centering
  \includegraphics[width=\linewidth]{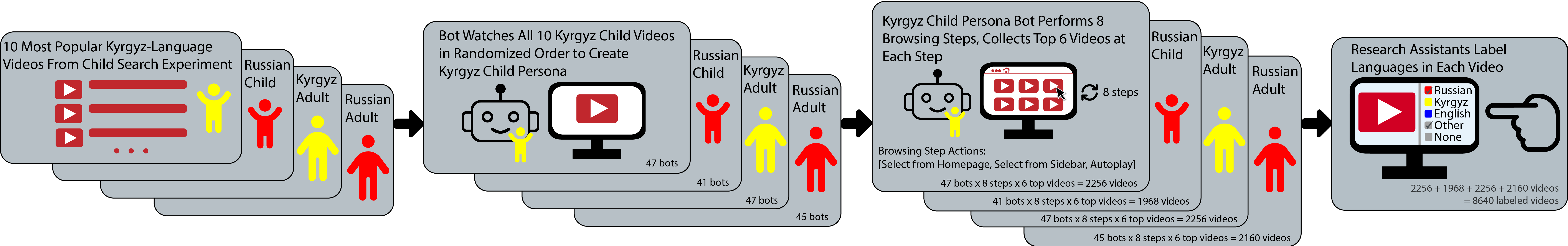}
  \caption{Steps performed in Experiment 3(a) by the browsing bots. Four personas are evaluated: Kyrgyz child, Russian child, Kyrgyz adult, Russian adult.}
  \Description{}
	\label{fig:rq3a_experiment}
\end{figure}

Our final experiment aimed to describe the recommendations that children and adults would encounter when browsing YouTube in Bishkek.

\subsubsection{Establishing Language- and Age-Based Personalization}

To evaluate the impact of language- and age-based personalization on recommendations, our sock-puppets establish personas by building a watch history. YouTube developers have specified that the recommendation algorithm incorporates the list of videos a user has previously watched \cite{covington2016}, with browser cookies providing tracking even when a user is not logged into an account \cite{haroon2022youtube}. Prior algorithmic audits of YouTube did not find a statistically significant impact on recommendations from specifying age directly through account settings \cite{hussein2020}, so we do not use Google accounts to signal demographic information to the personalization algorithm. Instead, we create a set of videos associated with a particular persona, and the sock-puppet watches these videos in sequence. 

Since Kyrgyzstan is not a supported region for the YouTube API or YouTube Kids, we compile videos associated with child and adult interests from the results of our search term experiments. An author fluent in Russian and Kyrgyz compiled four lists of videos (the top 10 Kyrgyz-language results returned for child-interest search terms, the top 10 Russian-language child-interest results, the top 10 Kyrgyz-language adult-interest results, and the top 10 Russian-language adult-interest results). Another author with language proficiency confirmed these lists. 

To trigger language- and age-based personalization, the sock-puppet watched all ten videos from a persona category (Kyrgyz child, Russian child, Kyrgyz adult, or Russian adult). Each sock-puppet opened the video URLs in a randomized order, and, following \cite{merrer2023}, each video was viewed for a randomized duration, with a mean of five minutes $\pm 30$ seconds, or the length of the video if shorter. Though algorithmic auditing studies have used varying numbers of videos to build personas \cite{srba2023, schellingerhout2023, hussein2020, merrer2023}, prior work has found strong personalization effects after viewing just three videos \cite{merrer2023}. We opted to increase the number of videos watched, as the algorithmic auditing literature has not yet tested personalization for age or location in our study context. While further experiments could test whether viewing additional seed videos would trigger stronger personalization, our ethnographic observations indicate that our current approach already signals an unusually strong user preference. Participants did not typically watch a large predefined set of videos; instead, their viewing behavior was dynamic and guided by recommendations.

\subsubsection{Browsing Behavior Experimental Setup}

We first test the recommendations made for four personas: children interested in Kyrgyz-language content, children interested in Russian-language content, adults interested in Kyrgyz-language content, and adults interested in Russian-language content. Each sock-puppet opens the YouTube platform, watches ten videos from the list corresponding to its assigned persona, then performs a random walk for eight steps. At each step, the sock-puppet either selects a recommendation or returns to the homepage. We record the top six recommendations encountered at each step. We executed and analyzed 47 trials for the Kyrgyz child persona, 41 for the Russian child persona, 47 for Kyrgyz adult persona, and 45 for the Russian adult persona\footnote{The numbers of trials are uneven, because some videos were deleted or made private in the time span between data collection and labeling. We chose to exclude trials with missing videos.}. This experiment creates a watch history to strongly signal interest in a type of content, then observes the recommendations encountered when browsing the platform. 

\begin{figure}[h!]
  \centering
  \includegraphics[width=\linewidth]{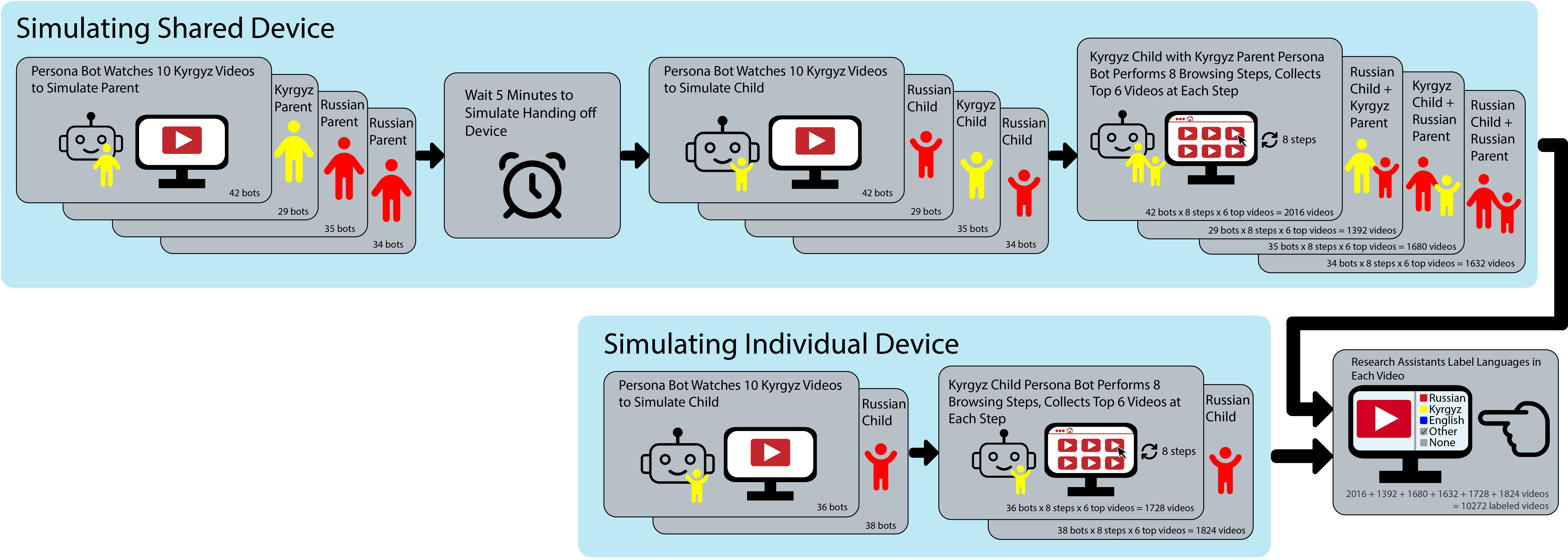}
  \caption{Steps performed in Experiment 3(b) by the browsing bots. Four device sharing arrangements are evaluated: Kyrgyz child and Kyrgyz parent, Russian child and Kyrgyz parent, Kyrgyz child and Russian parent, Russian child and Russian parent. Two individual device arrangements are evaluated: Kyrgyz child unshared, Russian child unshared.}
  \Description{}
	\label{fig:rq3b_experiment}
\end{figure}

Further, we investigated whether sharing a device would have an impact on recommendations that users received. Our ethnographic data documented many devices that were shared among family members, such as smart TVs and phones, rather than exclusively controlled by a single person. To simulate an adult caregiver handing off a device to their child, our sock-puppets first adopted an adult persona by watching ten videos, then paused for five minutes, then adopted a child persona by watching another ten videos. The sock-puppet then performed a random walk for eight steps. We executed and analyzed 36 trials of a Kyrgyz child persona on an individual device, 38 trials of a Russian child persona on an individual device, 29 trials of a device shared between a Kyrgyz adult persona and Russian child persona, 42 trials of a device shared between a Kyrgyz adult persona and a Kyrgyz child persona, 34 trials of a device shared between a Russian adult persona and Russian child persona, and 35 trials of a device shared between a Russian adult persona and Kyrgyz child persona\footnote{As with our other experimental setup, the number of trials is uneven because we opted to exclude trials with missing videos}. 

\subsection{Labeling Video Language}
\label{labeling}

Undergraduate research assistants from a university in Bishkek with advanced proficiency in Russian, Kyrgyz, and English hand-labeled the languages present in each video gathered during our experiments. They were compensated \$0.09 per video, labeling two videos per minute on average, for an hourly wage of \$10.80 USD. To facilitate this task, we designed a custom website that displayed an embedded video above a series of language selection buttons (Kyrgyz, Russian, English, Other, No spoken or written language). The buttons provided binary labels---a particular language was either present in a video or not. Research assistants could select multiple buttons to indicate that a video contained multiple languages. Two research assistants independently assessed each video. If labels conflicted, a third research assistant served as a tie-breaker. We recorded the majority decision as the final label for each video.

There was high agreement after the initial round of coding for the 10,741 unique videos\footnote{We collected a total of 26,016 video observations across all our experiments, but some videos were observed multiple times. The total number of unique videos is lower than the total video observations because we only label each video once.} collected during our experiments (Krippendorf's alpha of 0.86). The authors reviewed twenty conflicting labels and found that most disagreements were attributable to line-drawing issues rather than carelessness. For instance, research assistants disagreed on whether English was present in a ``Tom and Jerry'' cartoon, as an English-language title card was briefly displayed, but the animal characters did not speak during the episode. Though time-consuming, this approach produced high-quality labels.

We initially explored using machine classification to label our video dataset but decided against this approach. The only publicly available models that supported Kyrgyz could solely classify text (e.g., fastText), and there were no models for transcribing Kyrgyz audio. Further, the text classifiers could not signal whether multiple languages were present in a text sample. We considered using text classification on the video metadata but found instances where the video title and description did not reliably indicate the video's content (e.g., YouTube automatically translated the titles of popular English videos into Russian). At the time of this experiment, popular public machine learning models for classifying languages of audio and text were insufficient for our study purposes; hand-labeling was necessary to identify low-resource languages like Kyrgyz.

\section{Algorithmic Auditing Results}
\label{audres}

In this section, we report the results of a series of experiments designed to answer each of our research questions. To analyze experimental results, we conducted quantitative analysis. We built statistical models to study the association between the number of recommended YouTube videos in a specific labeled language (Kyrgyz, Russian, English, Other, or No language) and different variables corresponding to the experimental conditions. In particular, we used binomial regression where the dependent variable was the natural logarithm of the odds of observing a video with a specific language. Then, we used the models to estimate the probability (computed from the natural log odds) that a recommended video will be in a labeled language, and performed pairwise comparison with Bonferroni correction to test our null hypotheses that there is no difference between the probabilities of labeled languages. We report odds ratios, where values farther away from 1 indicate a stronger effect, and the corresponding p-values that indicate whether we can reject our null hypotheses.
\enlargethispage{2\baselineskip} 

\subsection{RQ1) What content does YouTube present to new users across Kyrgyzstan?}
\label{RQ1}

This location experiment measured the impact of geographic location on the videos that YouTube recommended to new users in Kyrgyzstan. To analyze our data, we used a binomial regression to study the association between location and number of videos on the homepage in a specific labeled language: 

\begin{equation}
    (video\ count_{labeled\ language},\ video\ count_{total}) \sim labeled\ language \cdot location
    \label{eq:location}
\end{equation}

\subsubsection{In the mixed-language capital city of Bishkek, new users receive about as many Kyrgyz-language as Russian-language recommendations.}

In Bishkek, prior to user personalization, the YouTube algorithm predominantly recommends Kyrgyz and Russian videos, with fewer videos in English (odds ratio=1.91, 95\% CI=[1.62, 2.26], $p <.001$ for both) or other languages (odds ratio=1.55, 95\% CI=[1.32, 1.81],  $p <.001$ for both). Our tests could not find a statistically significant difference between YouTube recommendations of Russian and Kyrgyz-language videos on the platform homepage for new users (odds ratio = 0.998, 95\% CI=[0.86, 1.15], $p=0.9788$) (see Figure \ref{fig:location}).

\subsubsection{Even in Kyrgyz-speaking regions of the country, new users receive high numbers of Russian-language recommendations.}

\begin{figure}[H]
  \centering
  \includegraphics[width=.6\linewidth]{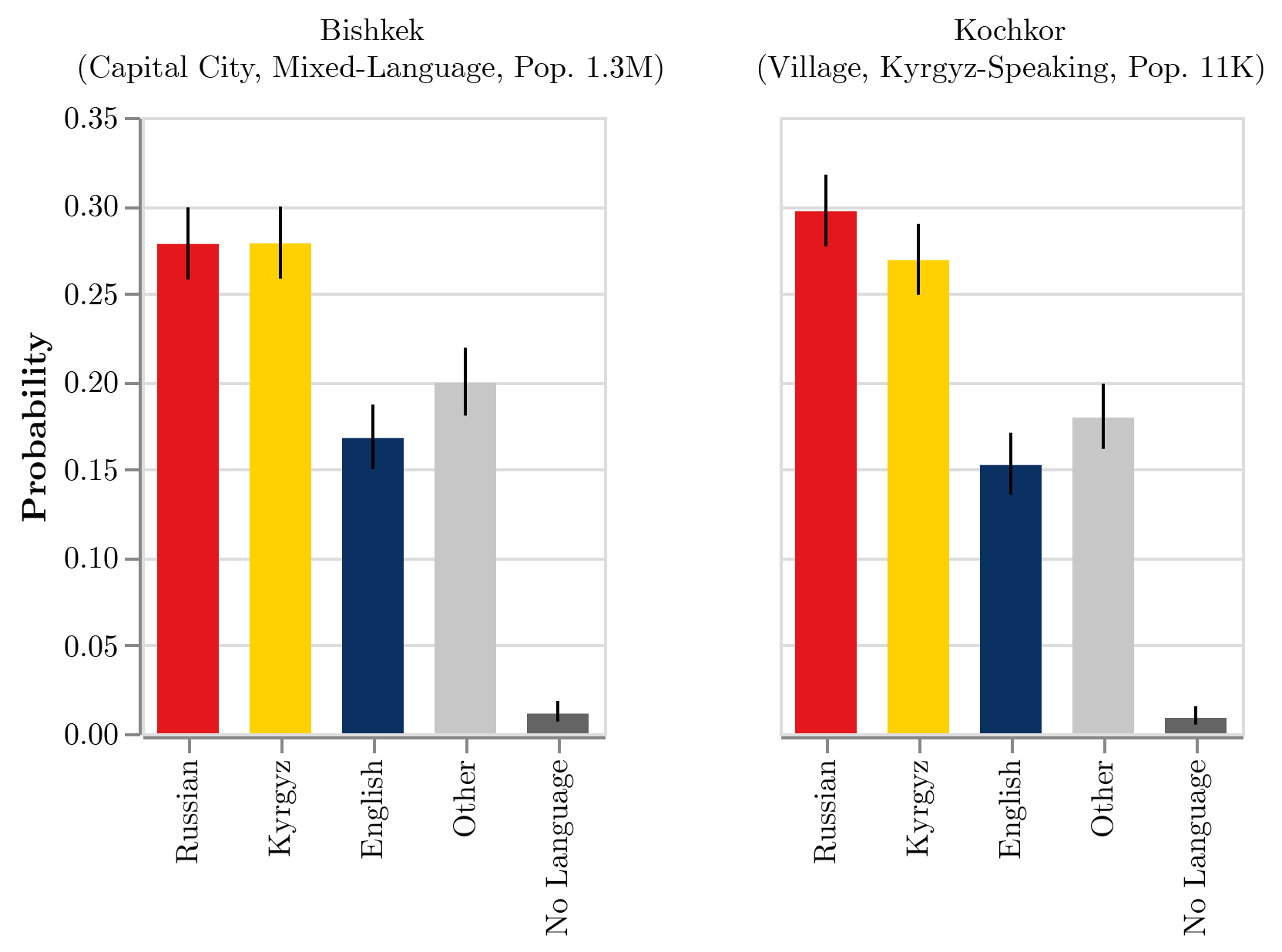}
  \caption{In the mixed-language capital of Bishkek, Kyrgyz and Russian videos appear on the homepage with similar probabilities. For the Kyrgyz-speaking village of Kochkor, Russian videos are slightly prioritized over Kyrgyz videos. Model-estimated probability of a video being recommended in a specific labeled language across Bishkek and Kochkor. The error bars show a 95\% confidence interval.}
  \Description{Bar chart of the model-estimated probability of a video being recommended in a specific labeled language (Russian, Kyrgyz, English, Other, or No Langauge) across Bishkek and Kochkor.}
	\label{fig:location}
\end{figure}

Conforming with our expectations, YouTube does not present new users in Kochkor with predominantly Kyrgyz recommendations (see Figure \ref{fig:location}). Despite YouTube's use of geolocation to personalize recommendations \cite{covington2016}, users in this Kyrgyz-speaking village receive slightly more Russian- than Kyrgyz-language recommendations  (odds ratio=1.15, 95\% CI=[1.00, 1.32], p=0.0566). Our tests could not find a statistically significant difference between the probabilities of Russian-language (odds ratio=1.10, 95\% CI=[0.95, 1.23], $p=0.2040$) and Kyrgyz-language  (odds ratio=0.95, 95\% CI=[0.82, 1.10], $p=0.5141$) videos in Kochkor and Bishkek. This suggests that the YouTube recommendation algorithm does not reflect the local linguistic context in other regions of Kyrgyzstan, potentially explaining the anecdotal accounts in our ethnography of children spontaneously speaking Russian in remote villages.

\subsection{RQ 2) What content does the YouTube search algorithm prioritize in Kyrgyzstan?}

Our next experiment described how the YouTube search algorithm responded to different queries. We first analyzed the results returned for the original terms collected from Google Trends. Since the original child-interest terms were more frequently written in Russian, we next explored whether the same trends persisted after translating the terms into Kyrgyz. Finally, we interrogated a potential mechanism for the prioritization of Russian by assessing the treatment of bivalent terms---terms that are written the same in Kyrgyz and Russian, such as transliterations of international pop culture terms---to determine which language the algorithm favors when presented with ambiguous queries.

\begin{figure}[H]
  \centering
  \begin{subfigure}{.9\textwidth}
  \includegraphics[width=\linewidth]{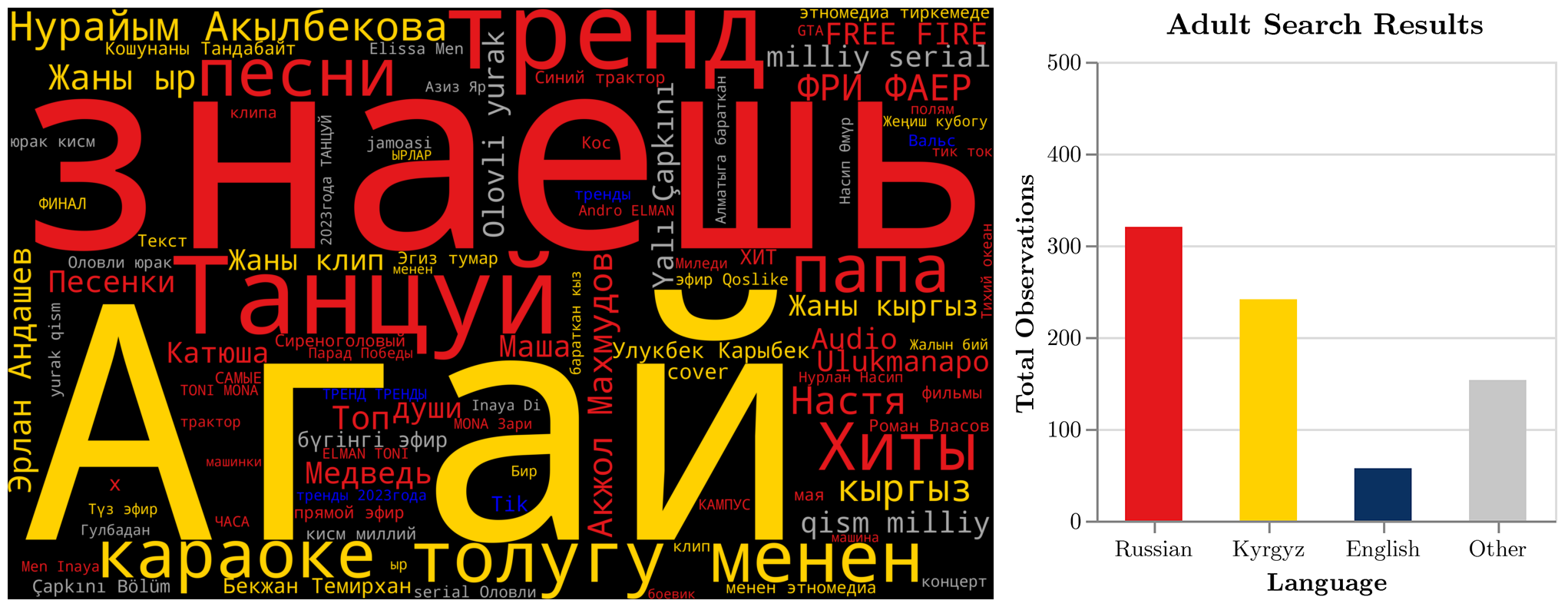}
  \caption{}
  \label{fig:adult}
\end{subfigure}
\begin{subfigure}{.9\textwidth}
  \includegraphics[width=\linewidth]{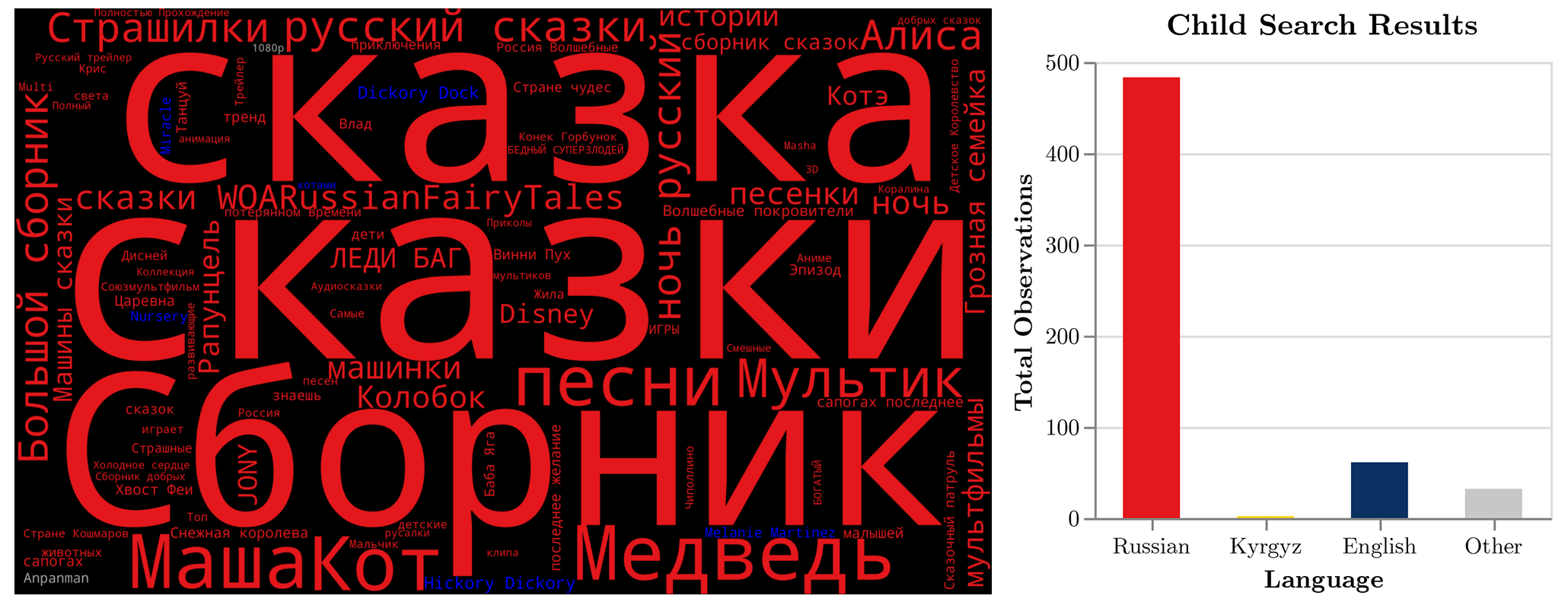}
  \caption{}
  \label{fig:child}
\end{subfigure}
  \caption{Word cloud of video titles returned for (a) adult-interest and (b) child-interest original Google search terms. Phrases that appear in Russian-language videos are colored in red, Kyrgyz in yellow, English in blue, and other or uncertain terms in gray. Bar graph provides raw counts of results returned in Russian, Kyrgyz, English, or Other languages for the 59 child-interest and 85 adult-interest original search terms.}
  \Description{}
	\label{fig:wordcloud}
\end{figure}

\subsubsection{Kyrgyz children's current search practices return predominantly Russian results.}

We first investigated the original search terms. In our ethnography, we observed many Kyrgyz children watching videos that they had located using the search function; this analysis explored the content their existing search practices surfaced. We expected some difference due to the language of original search terms: the majority of child-interest terms were written in Russian (71.1\%), compared to a minority of adult-interest terms (30.6\%). However, as we encountered several videos with mixed-language titles (e.g., Soviet-era cartoons with Kyrgyz voice-overs), we were unsure how the language of the search term impacted the language of the results. Yet, when our sock-puppets searched the 59 child-interest search terms taken directly from Google Trends, the YouTube search algorithm returned predominantly Russian-language results (see Figure \ref{fig:child}). These results contrast with the adult-interest terms, which returned results in a mix of languages (see Figure \ref{fig:adult}; as 84\% of ``Other'' observations were in Kazakh or Uzbek, Central Asian languages appeared in the majority of search results). This outcome strongly suggests that Kyrgyz children's current search practices will produce predominantly Russian-language search results.

\subsubsection{Even after translating terms into Kyrgyz, child-interest searches receive more Russian results than adult-interest searches.}

As more of the original child-interest terms from Google Trends were written in Russian than in Kyrgyz, we next aimed to determine whether the Russian tilt of our preliminary analysis would be corrected by balancing the number of searches issued in Russian and Kyrgyz. We compared the languages of the results returned for child-interest and adult-interest searches across search strategies, evaluating equal numbers of searches in Russian and Kyrgyz. We used a binomial regression to study the association between child- and adult-interest terms and the number of recommended videos in a specific labeled language:  
\begin{equation}
    (video\ count_{labeled\ language},\ video\ count_{total}) \sim labeled\ language \cdot age
    \label{eq:age}
\end{equation}

\begin{figure}[H]
  \centering
  \includegraphics[width=.6\linewidth]{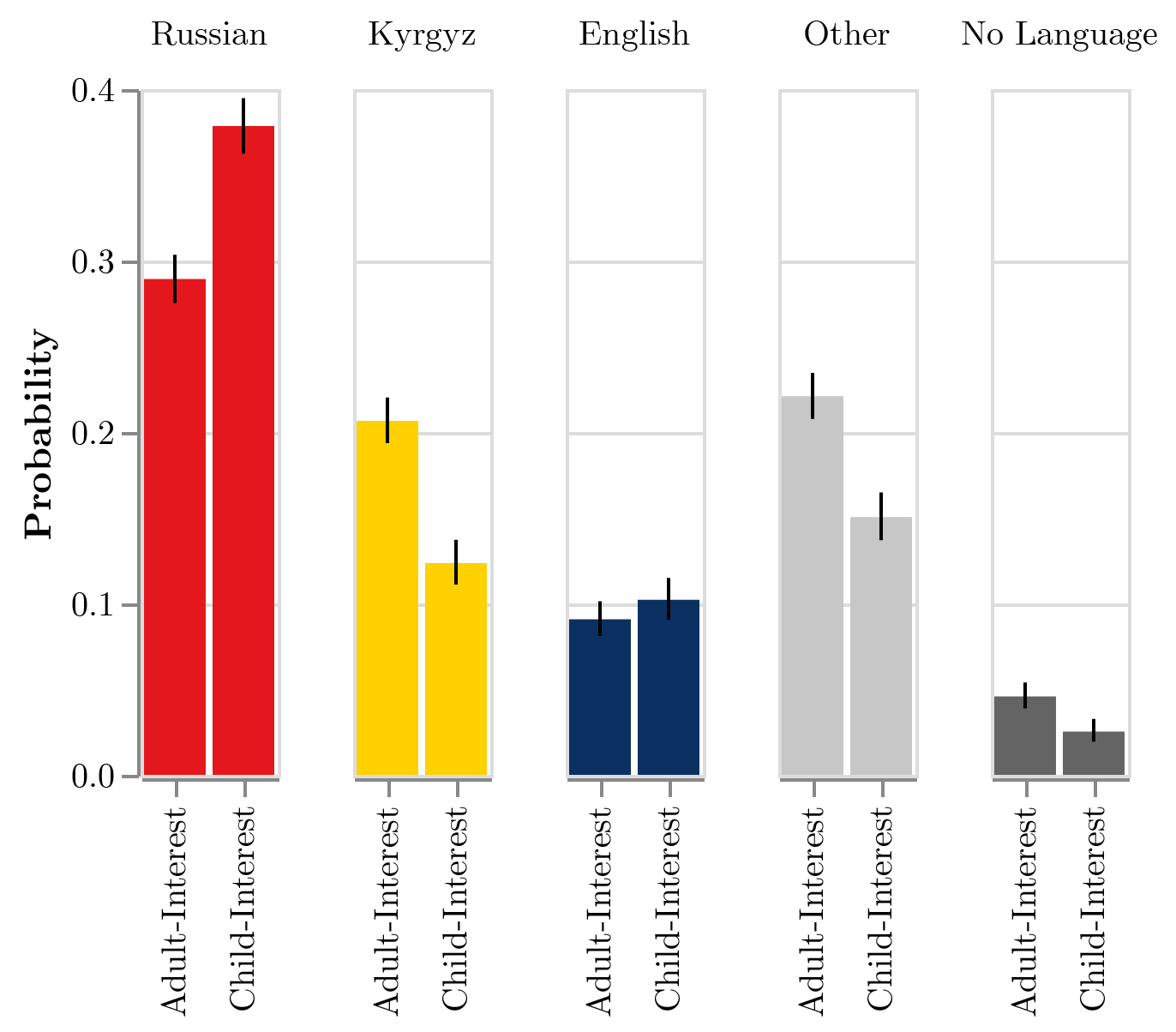}
  \caption{Child-interest search terms receive more Russian and fewer Kyrgyz recommendations than adult-interest terms across tested conditions. Model-estimated probability of a video being recommended in a specific labeled language across child- and adult-interest terms. The error bars show a 95\% confidence interval.}
  \Description{Bar chart of the model-estimated probability of a video being recommended in a specific labeled language (Russian, Kyrgyz, English, Other, or No Language) across child and adult-interest terms.}
	\label{fig:age}
\end{figure}

Across all search strategies, searching for a child-interest concept led to more Russian (odds ratio = 1.51, 95\% CI=[1.37, 1.68], $p<0.001$), more English (odds ratio = 1.27, 95\% CI=[1.04, 1.54], p=0.02087), and fewer Kyrgyz results (odds ratio = 0.12, 95\% CI =[0.07, 0.20], $p<0.001$) than adult-interest searches (see Figure \ref{fig:age}). This suggests that the YouTube algorithm recommends more Russian content in response to child-interest searches in Kyrgyzstan. These results support the beliefs of Kyrgyz caregivers who worried that their children received more Russian video recommendations than they did.

\subsubsection{Searches in Russian return more Russian results, searches in Kyrgyz return more Kyrgyz results, but bivalent searches, which are written the same in Russian and Kyrgyz, return more Russian results.}

We compared the languages of the results returned for 92 univalent Russian (i.e., words that appear in the Russian lexicon and not in Kyrgyz), 92 univalent Kyrgyz, and 52 bivalent terms (i.e., words that appear the same in Russian and Kyrgyz lexicons). To understand the content presented to children, it is important to investigate how the recommendation algorithm treats bivalent terms. Bivalent terms include transliterations of international pop culture terms that often appeal to children, e.g.,  TikTok is written as ``\foreignlanguage{russian}{тик ток}'' in both Kyrgyz and Russian. 


We used a binomial regression to study the association between the $valence$ of a search term (univalent Russian, univalent Kyrgyz, bivalent) and the number of recommended videos in a specific $labeled\ language$: 

\begin{equation}
    (video\ count_{labeled\ language},\ video\ count_{total}) \sim labeled\ language \cdot valence
    \label{eq:valence}
\end{equation}

Univalent Russian searches returned more videos in Russian than in any other language (all $p < 0.001$), and univalent Kyrgyz searches returned more videos in Kyrgyz than in any other language (all $p < 0.001$). 

\begin{figure}[H]
  \centering
  \includegraphics[width=.6\linewidth]{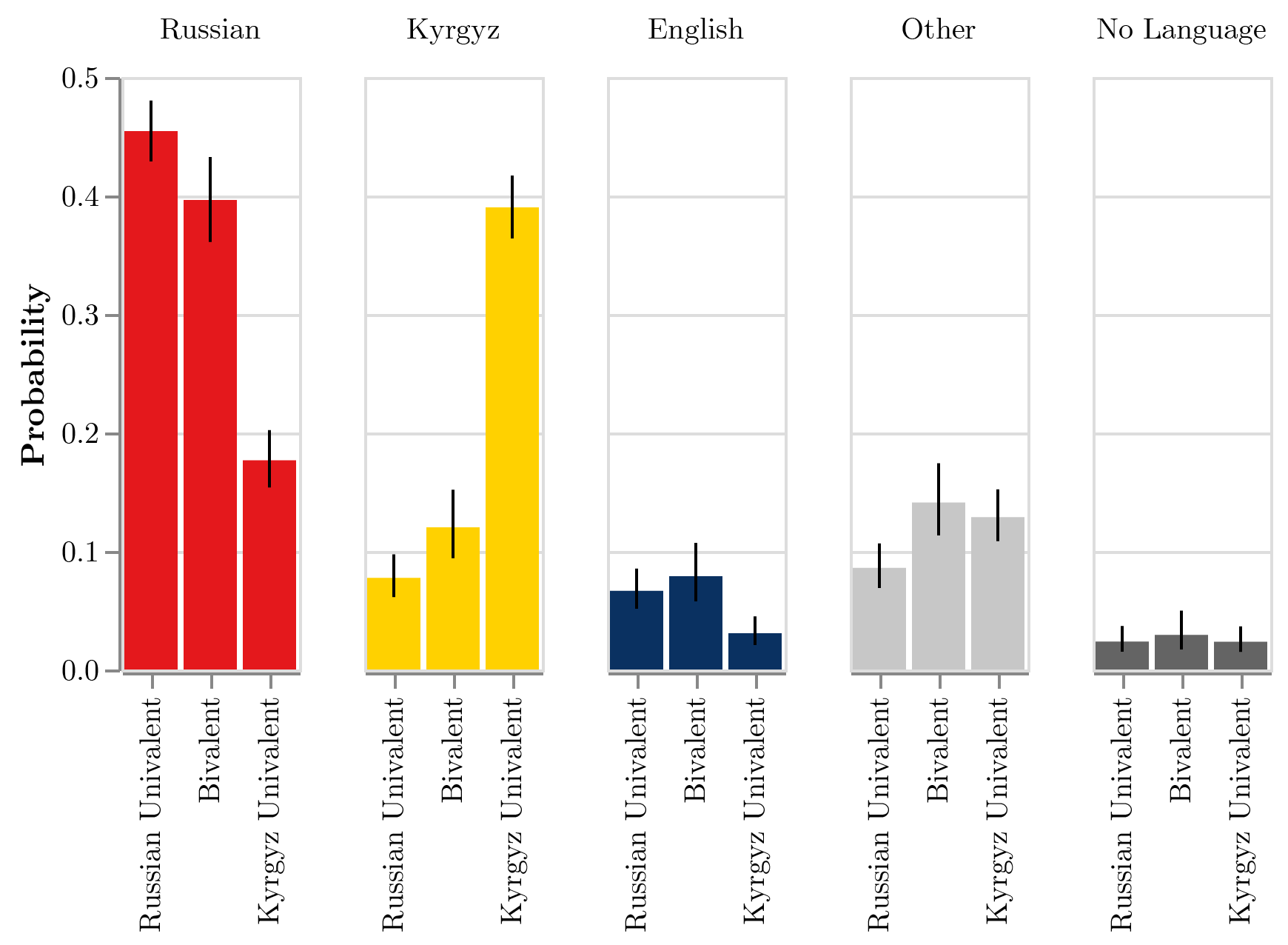}
  \caption{Bivalent terms return more Russian than fewer Kyrgyz recommendations, suggesting that the search algorithm does not treat bivalent terms as if they were written in Kyrgyz. Model-estimated probability of a video being recommended in a certain language across bivalent, Russian univalent, and Kyrgyz univalent interest terms. The error bars show a 95\% confidence interval.}
  \Description{Bar chart of the model-estimated probability of a video being recommended in a specific labeled language (Russian, Kyrgyz, English, Other, or No Langauge) across bivalent, Russian univalent, and Kyrgyz univalent interest terms.}
	\label{fig:valence}
\end{figure}

Bivalent search terms return significantly more Russian recommendations than Kyrgyz recommendations (odds ratio = 4.81, 95\% CI=[3.53, 6.56], $p < 0.001$; see Figure \ref{fig:valence}). This suggests that the YouTube search algorithm treats bivalent terms as if they were not written in Kyrgyz. It is unclear whether a user in Bishkek who searches for the bivalent terms ``\foreignlanguage{russian}{тик ток}''/TikTok, ``\foreignlanguage{russian}{Гарри Поттер}''/Harry Potter, or  ``\foreignlanguage{russian}{майнкрафт}''/Minecraft has a preference for Kyrgyz or Russian language content, but they will receive predominantly Russian recommendations regardless of their intent.

\subsection{RQ 3) How do the recommendations encountered by Kyrgyz children browsing the platform differ from those promoted to Kyrgyz adults?}

In this experiment, we describe the recommendations that children and adults encounter when browsing YouTube in Bishkek. We then explore whether sharing a device would produce different results. 

\begin{figure}[H]
  \centering
  \includegraphics[width=.8\linewidth]{images/watch.png}
  \caption{In contrast with other tested personas, watching a series of child-interest Kyrgyz videos did not lead to a Kyrgyz-language filter bubble during the browsing phase. Model-estimated probability of a video being recommended in a specific labeled language across Kyrgyz child, Kyrgyz adult, Russian child, and Russian adult personas over 8 browsing steps. The error shading shows a 95\% confidence interval.
}
  \Description{Line chart of the model-estimated probability of a video being recommended in a specific labeled language (Russian, Kyrgyz, English, Other, or No Langauge) across Kyrgyz child, Kyrgyz adult, Russian child, and Russian adult personas over 8 browsing steps.}
	\label{fig:age_browsing}
\end{figure}

\subsubsection{Unlike the other groups tested, signaling a strong interest in Kyrgyz child-interest content does not create a Kyrgyz-language filter bubble.}

We first compared the languages of the results returned for Kyrgyz child, Kyrgyz adult, Russian child, and Russian adult personas. We aimed to determine whether our personas would experience a filter bubble, which occurs when a user watches a certain type of content, and the search algorithm continues to promote similar content. 


We used a binomial regression to study the association  between child and adult personas who watched content in Russian or Kyrgyz and the number of recommended videos in a specific labeled language across 8 browsing steps: 

\begin{equation}
    (video\ count_{labeled\ language},\ video\ count_{total}) \sim labeled\ language \cdot age \cdot watch\ language \cdot step
    \label{eq:watch}
\end{equation}

We observed that some personas did experience a filter bubble: for instance, sock-puppets that watched Russian child-interest videos received predominantly Russian-language recommendations during all eight steps of the browsing phase ($p < 0.001$ for all languages at every step; see Figure \ref{fig:age_browsing}). Although the sock-puppets that watched Kyrgyz child-interest videos received more Kyrgyz-language videos at the initial browsing step ($p < .01$ compared to all other labeled languages at step 1), by the final browsing step, the YouTube algorithm recommended more English videos (odds ratio = 1.75, 95\% CI=[1.14, 2.69], $p=0.0105$) and content without spoken or written words (odds ratio = 1.60, 95\% CI=[1.04, 2.47], $p=0.0340$) than Kyrgyz-language videos. This is not a phenomenon unique to the language---sock-puppets that watched Kyrgyz adult-interest content received predominantly Kyrgyz-language recommendations across all steps in the browsing phase ($p < 0.001$  for all languages at every step), so it is possible to establish a Kyrgyz-language filter bubble.

Surprisingly, the YouTube algorithm recommended fewer Kyrgyz videos to children who have actively sought out and engaged with Kyrgyz-language content than to new users whose interests are unknown to the platform (compare Section \ref{RQ1}).

\subsubsection{Sharing a device within a family can help align recommendations with caregiver priorities.}

Next, we compared the languages of the results returned for Kyrgyz and Russian child personas who either had exclusive control over a device or shared it with a caregiver persona. We aimed to determine whether sharing a device with a Kyrgyz caregiver would increase the number of Kyrgyz-language recommendations relative to using a personal device. We used a binomial regression to study the association between Kyrgyz and Russian child personas who either had their own device or shared with a caregiver and the number of recommended videos in a specific labeled language across 8 browsing steps: 

\begin{equation}
    (video\ count_{labeled\ language},\ video\ count_{total}) \sim labeled\ language \cdot device\ ownership \cdot watch\ language \cdot step
    \label{eq:child_vs_adult}
\end{equation}

Our analysis indicates that children who share a device with an adult who watches Kyrgyz content will see more Kyrgyz-language recommendations than those who have their own device. This increase applies even when the children engage with Russian content ($p <0.001$ across all eight browsing steps). Though sock-puppets that watched Kyrgyz child-interest content experienced a significant decrease in the number of Kyrgyz videos recommended across the browsing session (step 8 vs. step 1: odds ratio = 0.33, 95\% CI=[0.21, 0.53], $p < 0.001$),  sock-puppets that emulated a Kyrgyz caregiver handing off their device to a Kyrgyz child did not see a significant drop (step 8 vs. step 1: $p = 0.0632$). Bots that emulated a Kyrgyz child using a device shared with a Kyrgyz adult encountered a Kyrgyz-language filter bubble when browsing the platform; Kyrgyz was more prevalent than English and Russian across all eight browsing steps ($p < 0.001$). 

To determine whether the Kyrgyz recommendations promoted on shared devices would appeal to children, we investigated the top 20 videos most frequently recommended on devices shared by children and adults with an interest in Kyrgyz-language content. Coding the content of these videos, we found that many suggestions were Kyrgyz-language movies. Interestingly, ``\foreignlanguage{russian}{фильм}''/movie was a mixed-interest term that we had filtered out from our test terms, as it appeared in both ``Children's Interests'' and ``All categories'' on Google Trends. When devices are shared within Kyrgyz-speaking families, it appears that the YouTube algorithm recommends Kyrgyz-language content that appeals to children and adults.

\begin{figure}[h]
  \centering
  \includegraphics[width=\linewidth]{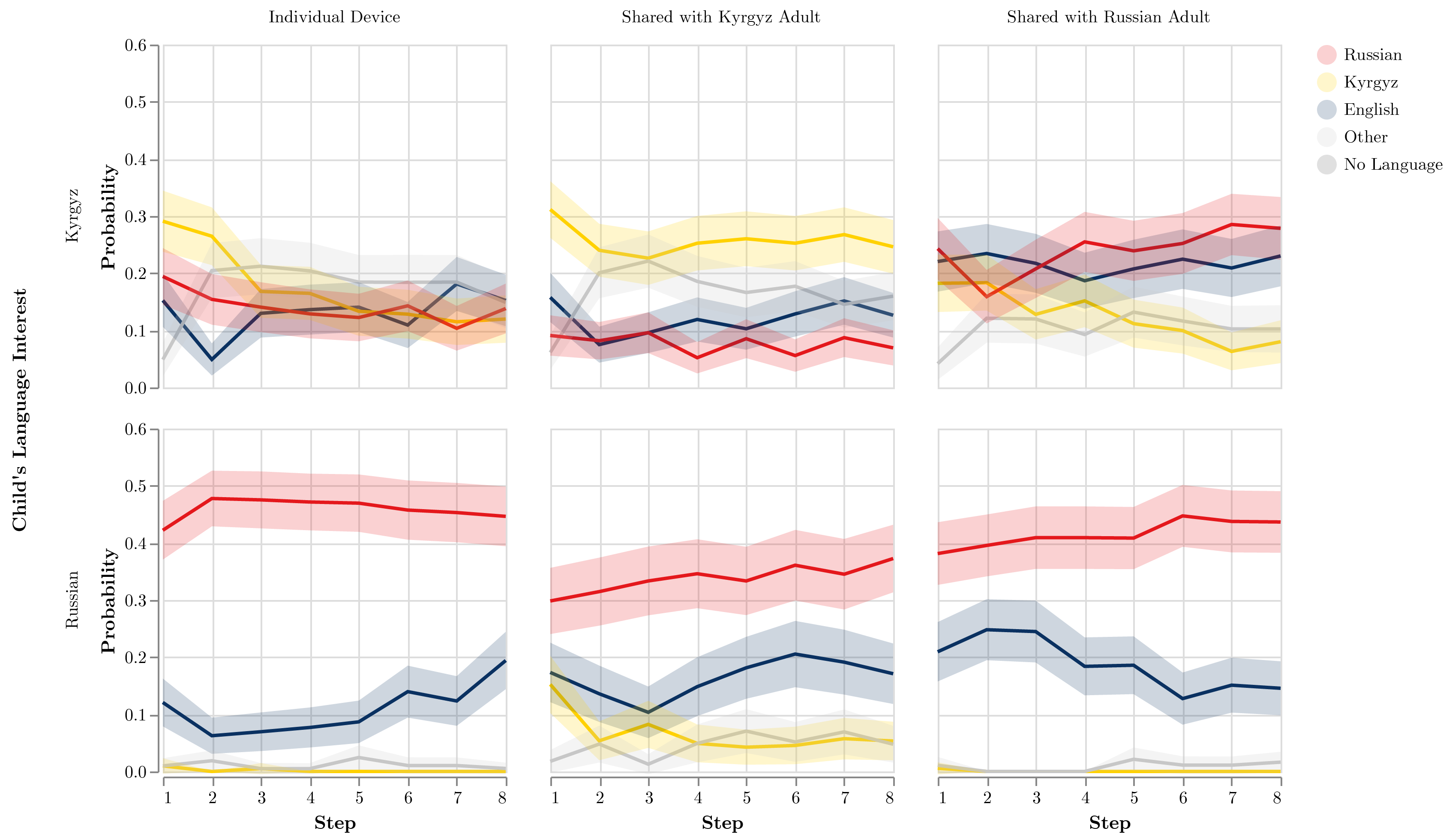}
  \caption{Unlike the persona emulating a Kyrgyz child using their own device, we observe a Kyrgyz-language filter bubble when a device is shared between a Kyrgyz child and Kyrgyz adult persona. Model-estimated probability of a video being recommended in a specific labeled language across Kyrgyz child, Russian child, Kyrgyz child + Kyrgyz adult, Kyrgyz child + Russian adult, Russian child + Kyrgyz adult, Russian child + Russian adult personas over 8 browsing steps. The error shading shows a 95\% confidence interval.
}
  \Description{Line chart of the model-estimated probability of a video being recommended in a specific labeled language (Russian, Kyrgyz, English, Other, or No Langauge)  across Kyrgyz child, Russian child, Kyrgyz child + Kyrgyz adult, Kyrgyz child + Russian adult, Russian child + Kyrgyz adult, Russian child + Russian adult personas over 8 browsing steps.}
	\label{fig:sharing}
\end{figure}

\subsection{Characterizing Colonial YouTube Recommendations in Kyrgyzstan}

In this section, we establish that YouTube's amplification of colonial content cannot be blamed on a dearth of Kyrgyz-language content, as we identify thousands of hours of child-interest videos. Additionally, we establish that the Russian-language content promoted to Kyrgyz children is not produced by Indigenous creators and explore how the recommendation algorithm also amplifies imperial racial hierarchies.

\subsubsection{Availability of Kyrgyz-language Children's Content}

The imbalance we observe is not due to YouTube exhausting its reserves of Kyrgyz-language children's videos when making recommendations; such content is plentiful. In the ethnography, caregivers often named two channels as sources of at Kyrgyz revitalization content (state-sponsored TV Channel Balastan and international aid project UNICEF Kyrgyzstan); we use the YouTube API to total the length of all videos posted on these channels and find over 1300 hours of child-interest content. A child who watches an hour of YouTube daily would need almost four years to make it through just those two channels. Further, this institutional revitalization content only amounted to 20\% of the Kyrgyz-language child-interest videos we collected during our study; a wide range of creators contribute to YouTube's Kyrgyz-language kids catalog. For instance, the 35th most viewed channel on the entire platform, surpassing large acts like American singer Taylor Swift and K-pop group BLACKPINK, belongs to Bishkek-based group D Billions (@dbillions) \cite{socialblade}. This group exclusively produces children's content, and their dedicated Kyrgyz-language channel (@dbillions\_kyrgyz) lists over 600 videos. Given this expanse of available content and ethnographic observations of children re-watching videos, it should be possible for a child to consume exclusively Kyrgyz-language content on YouTube. Notwithstanding the enormous efforts to provide content in their heritage language, Kyrgyz children are inundated with Russian-language recommendations.

\subsubsection{Recommended Russian-language Videos Reinforce Imperial Racial Hierarchies}
\label{race}

Our findings reveal that the YouTube recommendation algorithm prioritizes Russian-language content for children in Bishkek, confirming the concerns raised by Kyrgyz caregivers in our ethnography. However, linguistic ideologies are not the only colonial logics that structure the content recommended to children.

Supplementing our language analysis, we also assessed whether the use of Russian in YouTube's recommended videos is racialized. As it was possible that Kyrgyz creators could be posting Russian-language videos, the Russian-language content may have had Indigenous origins. Our Kyrgyz research assistants labeled the perceived ethnicity of individuals in our video dataset, in addition to the languages written and spoken. A Kyrgyz author selected the following relevant ethnic categories in Kyrgyzstan: Kyrgyz, Other Central Asian, Person from the Caucasus, Slavic, and Other Ethnicities. In our dataset, we identify 1378 YouTube videos with ethnically Kyrgyz people; most spoke Kyrgyz (90.3\% of 1378 videos), and a few spoke Russian (8.6\%). The Russian-language videos we analyze rarely feature ethnically Kyrgyz people (2.7\% of 4424 Russian-language videos). We conclude that the recommended Russian-language content typically does not have Indigenous origins, instead imposing imperial perspectives.

When Kyrgyz children use the YouTube platform, the Russian-language videos that are recommended are unlikely to feature ethnically Kyrgyz people. Rather, children see Slavic content creators or cartoons with Slavic features, such as Masha from ``Masha and the Bear'' who dresses in traditional Russian folk costume. This spotlight on Slavic representations can further entrench colonial scales of difference, diminishing the visibility and cultural significance of Kyrgyz people within digital media.

\section{Discussion \& Future Work}
\label{sec:discuss}

Contrary to the optimistic belief that platforms will help Indigenous languages flourish and reach young audiences \cite{dale2022case, davis2018talking, moshnikov2022use, wang2022minority}, we argue that YouTube is a homogenizing force that amplifies imperial content. Our study shows that YouTube's recommender system prioritizes Russian-language children's content in Kyrgyzstan, despite revitalization initiatives like UNICEF Kyrgyzstan providing ample material in Kyrgyz. Whether children are searching for terms or browsing recommendations, they will encounter more Russian than Kyrgyz content---even after they strongly signal a preference for Kyrgyz. While there is a possibility that this skew may be explained by referencing the relatively fewer number of Kyrgyz- compared to Russian-language videos, there is strong counter-evidence: if recommendations were based upon the supply of content, we would expect far more videos to be in English, the most popular upload language on YouTube \cite{spicer}. As YouTube was developed primarily in Silicon Valley but propagates Russian content in a former Soviet republic, we show that platforms created by one power can reinforce the colonial hierarchies of a different imperial core. 

These findings are critical because the platform’s colonial prioritization actively shapes the identities and offline practices of Indigenous youth. By 'bundling' \cite{keane2003semiotics} children’s interests with the imperial language, the platform socializes youth to view Russian as the default language of entertainment and technology and to view Kyrgyz as uninteresting. The result is language shift: children are speaking more Russian and adopting linguistic features from YouTube that diverge from their parents' Russian in Bishkek, such as a phonologically more ``Moscow'' dialect and the absence of code-switching. Platform recommendations do not merely reflect existing youth preferences; recommendation architectures can actively erode Indigenous cultural practice online and offline.

\subsection{Platforms: Design to Support Indigenous Self-Determination}

It is impossible to fully detangle the YouTube recommender system from issues of power; ranking is itself a colonial logic \cite{bowker1999sorting}. A private firm organizes and indexes all the content on the platform, creating and imposing scales of relevance for a global set of users \cite{zuboff, noble2018}. They optimize algorithms to generate profit \cite{zuboff, noble2018}, not prioritize content desired or valued by Indigenous people.

Though our algorithmic audit detects amplification of non-Indigenous content, this is not an inevitable outcome of platform design. The language imbalance we observe is a product of the YouTube recommendation algorithm, which combines many factors (such as video freshness, user geographic location \cite{covington2016}, and sponsorship \cite{ytads}) and computational approaches (such as recurrent neural networks \cite{beutel18}). It may be possible in future work to identify the exact mechanisms that prioritize Russian content in Kyrgyzstan; some candidates might include the proportion of videos available in each language or popularity metrics. However, providing a causal mechanism or technical explanation does not justify the observed effects. As YouTube itself acknowledges, ``[W]hile clicks, watchtime, user surveys, shares, likes and dislikes are important signals that inform our system, they can be overruled by our commitment to meeting our responsibility to the YouTube community and to society'' \cite{goodrow2021}. 

YouTube and other platforms could make design choices that support Indigenous self-determination. For instance, to counter the amplification of Russian-language content, YouTube could introduce filters that allow users to set language preferences for their recommendations. This would give users the agency to curate their own online linguistic environments. Platforms could also directly mitigate the mechanisms of colonial amplification by refining how geographic signals are utilized. For instance, rather than defaulting to imperial languages for bivalent search terms, systems could use geographic data to prioritize Indigenous-language results, ensuring that users in Kyrgyzstan are met with heritage content by default (RQ2). YouTube has historically acted to protect children, implementing technical interventions that ``limit repeated recommendations'' of videos that push problematic body ideals or depict verbal aggression \cite{youthtube}. A similar approach could limit repeated recommendations of Russian-language content for Kyrgyz children, or the platform could adopt proposals from fairness researchers that address disparities among geographic groups \cite{gomez2024} or strengthen cultural citizenship \cite{ferraro2024}. Platforms should consult with Indigenous communities and prioritize their desired changes.

However, it may be long before YouTube changes its recommendation algorithm to counter the observed colonial effects, as private tech firms often focus their expenditures and efforts in Western contexts and neglect problems elsewhere \cite{reed2021, nigatu2024, silva2020}. Thus, in the remainder of this section, we consider how independent researchers can act with Indigenous communities to identify and address problematic algorithmic behavior.

\subsection{Users: Engage in Collective Protest and Employ Resistance Strategies}

Accomplishing broader changes to platform algorithms will likely require coordinated, collective protest. Our study points to some potential demands for popular campaigns, such as allowing users to filter recommendations by language. Users can engage in advocacy to put pressure on platforms to realign their designs with Indigenous interests.

Prior to platform reform, users can also develop a toolkit of resistance strategies to reshape their online spaces. Our results identify several methods for algorithmic contestation. First, we suggest that caregivers abstain from purchasing separate devices for family members; sharing devices leverages the established Kyrgyz-language watch history of caregivers to create Kyrgyz-language filter bubbles (RQ3). Second, caregivers can use translation as resistance: by translating a child's Russian-language request into Kyrgyz before submitting a search, they can ensure the algorithm surfaces more heritage content (RQ2). Finally, users can bypass the recommendation system altogether by creating and sharing playlists of Kyrgyz-language content. By testing and exchanging strategies, users can develop a set of practices that help sustain their cultural heritage in a biased digital landscape.

\subsection{HCI Researchers: Broaden Scope of Algorithmic Audits and Support Indigenous Digital Sovereignty}

As an initial step for HCI research, we echo calls to expand the current scope of algorithmic auditing efforts \cite{metaxa2021, urman2024, bandy2021} so that concerns from the global core are not prioritized over those of the periphery. Though we did not use racial classification as an experimental variable, a post-hoc analysis of our video dataset confirms the content YouTube recommends to Kyrgyz children features more Slavic than Central Asian people (Section \ref{race}). Future work could investigate racial and ethnic classification systems that depart from U.S. racialization schemes. Investigations in other former colonies or Indigenous communities could help address different concerns about digital algorithms. Other Indigenous groups also worry that technology is erasing their language \cite{millan2024}, but there are likely additional distressing issues that have not received adequate attention.

To identify these harms, HCI researchers must ensure that investigations are community-directed. When designing studies, researchers can impose paradigms that are not portable to other contexts. For instance, we attempted to replicate this audit in the formerly colonized regions of Haiti, Algeria, and Ukraine. In Algeria, however, we found that YouTube most frequently recommended Quranic \textit{surah} recitations, not French-language content as we anticipated, suggesting that popular use of the platform diverged from the more entertainment-oriented Kyrgyz practices. It is inappropriate to assume that the concerns and context of one Indigenous group carries over to others \cite{noe2024}. Thus, we recommend that researchers draw upon qualitative research and/or partner with communities before designing algorithmic audits in Indigenous contexts to ensure that the variables measured reflect the community’s lived anxieties.


Finally, HCI scholars must move from documenting harm to promoting digital sovereignty. Indigenous communities should have power over the algorithms that structure their online worlds, rather than private platforms dictating how their youth encounter information. Future work could explore how to elicit and incorporate specific Indigenous data practices and values into algorithmic systems, drawing inspiration from existing decolonial frameworks such as Barrett et al.'s articulation of African data ethics from \cite{barrett25}. Algorithmic audits could play a role here---moving beyond descriptive studies---by articulating explicit technical demands and determining whether adequate concessions have been implemented. Decolonial thinking does not demand that Indigenous people refuse existing technology, but prompts us to consider how it can be appropriated to advance the self-determination of those at the periphery.

\section{Conclusion}

Scholars and technology companies tout online platforms as vehicles for revitalizing Indigenous cultural heritage. Indigenous communities remain skeptical of products developed in wealthy tech hubs, and voice concerns that platforms may actually imperil, rather than preserve, their languages, customs, and values. By combining ethnographic insights with a rigorous, community-informed algorithmic audit, we demonstrate that online platforms can reinscribe colonial hierarchies, rather than preserving Indigenous cultural heritage. First, we reviewed qualitative research concerning the cultural anxieties of a specific Indigenous community; we recognize that Indigenous peoples are not a monolith. We draw on our prior ethnographic work that identified deep-seated fears among Kyrgyz caregivers that YouTube encourages their children to speak Russian instead of Kyrgyz, their heritage language. Next, we constructed and conducted an algorithmic audit in conversation with this community. By tethering our inquiries to community interests, our approach reorients algorithmic auditing away from its typical Western-centric concerns and foregrounds a context shaped by colonial legacies. Our algorithmic audit results revealed that YouTube predominantly pushes non-Kyrgyz language content to Kyrgyz children---even when these users actively signal preferences for their heritage language. Attempts by children to make their identities legible to the algorithm were rebuffed; instead, the recommendation algorithm persistently constructed them as a Russian- or English-language audience. Reading our audit results alongside our ethnographic observations, the amplification of non-Kyrgyz material appears to encourage the offline uptake of colonial linguistic hierarchies---Kyrgyz children prefer to speak Russian, and YouTube promotes Russian videos that they quote and emulate, reinforcing their preference. Thus, the amplification of imperial content potentially accelerates the eroding of Indigenous practices offline. 

However, there are paths to challenge these algorithmically-determined outcomes and improve the visibility of Indigenous digital content. Through our algorithmic audit, we tested practical strategies for contesting algorithmic recommendations. We identified certain practices---such as sharing devices within families rather than giving children their own phones---that bring recommendations to align better with Kyrgyz preferences. Though algorithmic curation may sideline Indigenous content, we show there are possibilities for developing an end-user toolkit for resistance. With platform cooperation and the implementation of new features, such as language filters, it may be possible to rebuild systems so that they promote Indigenous content as a rule. Yet, such efforts may require significant institutional pressure, as the private firms that operate platforms tend to concentrate harm mitigation efforts in dominant language communities and are slow to respond to Indigenous populations.

Our findings warrant a broader, decolonial approach to HCI research and platform design. We emphasize that claims for Indigenous self-determination must extend into the digital domain; it is critical to foreground Indigenous priorities and perspectives when evaluating, contesting, and ultimately reshaping global platforms. Alongside ongoing decolonial computing projects that empower Indigenous communities to (co-)create new technologies, our work underscores the urgent need to address inequalities in existing technical systems that are adopted worldwide. Collaborating with communities can re-envision platforms that support Indigenous language, culture, and autonomy in a globalized world.

\begin{acks}

We are grateful to Snehal Prabhudesai, Jeff Bilik, Farnaz Jahanbakhsh, and our wonderful reviewers for providing comments on previous drafts of this paper. Abner Heredia Bustos from the University of Michigan's Consulting for Statistics Computing \& Analytics Research (CSCAR) provided assistance with our statistical analysis, though all faults are our own. We are indebted to Vicente Bobadilla Riquelme for research assistance and enthusiasm at early stages of this project. We benefited from opportunities to present versions of this work at the Algorithmic Fairness and Opacity Working Group (AFOG) at UC Berkeley; the Green Group at the University of Michigan; the Ling Lab at the University of Michigan; the Association for Slavic, East European, and Eurasian Studies; and the American Anthropological Association Annual Meeting. Finally, this work would not be possible without the families who participated in our ethnographic research. We thank you for your trust and knowledge.

\end{acks}

\bibliographystyle{ACM-Reference-Format}
\bibliography{bib}

\appendix





\section{Child Search Terms}
\label{appendix:childterms}

  \begin{filecontents*}{appendix_search_child_terms.csv}
Original Search Term,English Translation (Not Tested; Provided for Readability),Russian Translation,Kyrgyz Translation,Bivalent
\foreignlanguage{russian}{сказка},fairy tale,\foreignlanguage{russian}{сказка},\foreignlanguage{russian}{жомок},
\foreignlanguage{russian}{сказки},fairy tales,\foreignlanguage{russian}{сказки},\foreignlanguage{russian}{жомоктор},
\foreignlanguage{russian}{колобок},bun,\foreignlanguage{russian}{колобок},\foreignlanguage{russian}{көмөч нан},
\foreignlanguage{russian}{маша и медведь маша и медведь},Masha and the Bear Masha and the Bear,\foreignlanguage{russian}{Маша и Медведь Маша и Медведь},\foreignlanguage{russian}{Маша жана аюу Маша жана аюу},
\foreignlanguage{russian}{теремок},teremok,\foreignlanguage{russian}{теремок},\foreignlanguage{russian}{теремок},\checkmark
\foreignlanguage{russian}{царевна},tsarevna,\foreignlanguage{russian}{царевна},\foreignlanguage{russian}{принцесса},
\foreignlanguage{russian}{русалка},mermaid,\foreignlanguage{russian}{русалка},\foreignlanguage{russian}{суу периси},
\foreignlanguage{russian}{холодное сердце},Frozen (lit. Cold Heart),\foreignlanguage{russian}{Холодное Сердце},\foreignlanguage{russian}{Аяз кучагы},
\foreignlanguage{russian}{русские сказки},Russian fairy tales,\foreignlanguage{russian}{Русские сказки},\foreignlanguage{russian}{Орус жомоктору},
\foreignlanguage{russian}{мультики для мальчиков},cartoons for boys,\foreignlanguage{russian}{мультики для мальчиков},\foreignlanguage{russian}{балдар үчүн мультфильмдер},
\foreignlanguage{russian}{теремок тв},Teremok TV,\foreignlanguage{russian}{Теремок ТВ},\foreignlanguage{russian}{Теремок ТВ},\checkmark
\foreignlanguage{russian}{игры},games,\foreignlanguage{russian}{игры},\foreignlanguage{russian}{оюндар},
\foreignlanguage{russian}{рапунцель},Rapunzel,\foreignlanguage{russian}{Рапунцель},\foreignlanguage{russian}{Рапунцель},\checkmark
\foreignlanguage{russian}{кот в сапогах},Puss in Boots (lit. Cat in Boots),\foreignlanguage{russian}{Кот в сапогах},\foreignlanguage{russian}{Өтүкчөн мышык},
\foreignlanguage{russian}{кот в сапогах 2},Puss in Boots 2 (lit. Cat in Boots 2),\foreignlanguage{russian}{Кот в сапогах 2},\foreignlanguage{russian}{Өтүкчөн мышык 2},
\foreignlanguage{russian}{дети},kids,\foreignlanguage{russian}{дети},\foreignlanguage{russian}{жаш балдар},
\foreignlanguage{russian}{хвост феи},fairy tail,\foreignlanguage{russian}{хвост феи},\foreignlanguage{russian}{перинин куйругу},
\foreignlanguage{russian}{страшилки},horror stories,\foreignlanguage{russian}{страшилки},\foreignlanguage{russian}{коркунучтуу окуялар},
\foreignlanguage{russian}{гарри поттер},Harry Potter,\foreignlanguage{russian}{Гарри Поттер},\foreignlanguage{russian}{Гарри Поттер},\checkmark
\foreignlanguage{russian}{колобок сказка},bun fairy tale,\foreignlanguage{russian}{колобок сказка},\foreignlanguage{russian}{көмөч нан жомок},
\foreignlanguage{russian}{алиса},Alice,\foreignlanguage{russian}{Алиса},\foreignlanguage{russian}{Алиса},\checkmark
\foreignlanguage{russian}{сказка о потерянном времени},"""Tale about the Lost Time""",\foreignlanguage{russian}{сказка о потерянном времени},\foreignlanguage{russian}{жоголгон убакыт жомогу},
\foreignlanguage{russian}{чиполлино},Cipollino,\foreignlanguage{russian}{Чиполлино},\foreignlanguage{russian}{Чиполлино},\checkmark
\foreignlanguage{russian}{джони песни},Joni songs,\foreignlanguage{russian}{Джони песни},\foreignlanguage{russian}{Джони ырлары},
\foreignlanguage{russian}{машины сказки},Masha's tales,\foreignlanguage{russian}{машины сказки},\foreignlanguage{russian}{Машанын жомоктору},
\foreignlanguage{russian}{коралина в стране кошмаров},Coraline,\foreignlanguage{russian}{коралина в стране кошмаров},\foreignlanguage{russian}{Коралин},
\foreignlanguage{russian}{винни пух},Winnie the Pooh,\foreignlanguage{russian}{Винни Пух},\foreignlanguage{russian}{Винни Пух},\checkmark
\foreignlanguage{russian}{песни джони},songs Joni,\foreignlanguage{russian}{песни Джони},\foreignlanguage{russian}{ырлары Джони},
\foreignlanguage{russian}{маша и медведь 12 месяцев},Masha and the Bear 12 months,\foreignlanguage{russian}{Маша и медведь 12 месяцев},\foreignlanguage{russian}{Маша жана Аюу 12 ай},
\foreignlanguage{russian}{леди диана},Lady Diana,\foreignlanguage{russian}{Леди Диана},\foreignlanguage{russian}{Диана Айым},
\foreignlanguage{russian}{сказочный патруль},Fantasy Patrol,\foreignlanguage{russian}{Сказочный патруль},\foreignlanguage{russian}{Сыйкыр патруль},
\foreignlanguage{russian}{кошка},cat,\foreignlanguage{russian}{кошка},\foreignlanguage{russian}{мышык},
\foreignlanguage{russian}{a4},a4,\foreignlanguage{russian}{a4},\foreignlanguage{russian}{a4},\checkmark
\foreignlanguage{russian}{бартолито петушок},Bartolito cockerel,\foreignlanguage{russian}{Бартолито петушок},\foreignlanguage{russian}{Бартолито короз},
\foreignlanguage{russian}{тренды},trends,\foreignlanguage{russian}{тренды},\foreignlanguage{russian}{тренддер},
\foreignlanguage{russian}{кружит вайб},the vibe is circling,\foreignlanguage{russian}{кружит вайб},\foreignlanguage{russian}{вайб айланат},
\foreignlanguage{russian}{непета страшилки},nepeta horror stories,\foreignlanguage{russian}{непета страшилки},\foreignlanguage{russian}{непета коркунучтуу окуялар},
\foreignlanguage{russian}{влад а4},Vlad a4,\foreignlanguage{russian}{Влад а4},\foreignlanguage{russian}{Влад а4},\checkmark
\foreignlanguage{russian}{мальчик который обуздал ветер},the boy who harnessed the wind,\foreignlanguage{russian}{мальчик который обуздал ветер},\foreignlanguage{russian}{шамалды кармаган бала},
\foreignlanguage{russian}{hickory dickory dock},hickory dickory dock,\foreignlanguage{russian}{гикори дикори док},\foreignlanguage{russian}{гикори дикори док},\checkmark
\foreignlanguage{russian}{anpanman},anpanman,\foreignlanguage{russian}{анпанман },\foreignlanguage{russian}{анпанман },\checkmark
\foreignlanguage{russian}{страшные сказки},scary tales,\foreignlanguage{russian}{страшные сказки},\foreignlanguage{russian}{коркунучтуу жомоктор},
\foreignlanguage{russian}{челлендж},challenge ,\foreignlanguage{russian}{челлендж},\foreignlanguage{russian}{челлендж},\checkmark
\foreignlanguage{russian}{мультики для малышей},cartoons for kids,\foreignlanguage{russian}{мультики для малышей},\foreignlanguage{russian}{балдар үчүн мультфильмдер},
\foreignlanguage{russian}{пиноккио},pinocchio,\foreignlanguage{russian}{пиноккио},\foreignlanguage{russian}{пиноккио},\checkmark
\foreignlanguage{russian}{серая шейка},gray neck,\foreignlanguage{russian}{серая шейка},\foreignlanguage{russian}{боз моюн},
\foreignlanguage{russian}{конек горбунок},The Little Humpbacked Horse,\foreignlanguage{russian}{конек горбунок},\foreignlanguage{russian}{Кичинекей өркөч ат},
\foreignlanguage{russian}{miracle},miracle,\foreignlanguage{russian}{чудо},\foreignlanguage{russian}{керемет},
\foreignlanguage{russian}{melanie martinez},Melanie Martinez,\foreignlanguage{russian}{Мелани Мартинес},\foreignlanguage{russian}{Мелани Мартинес},\checkmark
\foreignlanguage{russian}{грозная семейка},formidable family,\foreignlanguage{russian}{грозная семейка},\foreignlanguage{russian}{каардуу үй-бүлө},
\foreignlanguage{russian}{машкины страшилки},Mashka's horror stories,\foreignlanguage{russian}{Машкины страшилки},\foreignlanguage{russian}{Машканын коркунучтуу окуялары},
\foreignlanguage{russian}{волшебные покровители},magical patrons,\foreignlanguage{russian}{волшебные покровители},\foreignlanguage{russian}{сыйкырдуу колдоочулар},
\foreignlanguage{russian}{кызыл раушан},red rose,\foreignlanguage{russian}{Красная роза},\foreignlanguage{russian}{кызыл раушан},
\foreignlanguage{russian}{влад и никита},Vlad and Nikita,\foreignlanguage{russian}{Влад и Никита},\foreignlanguage{russian}{Влад жана Никита},
\foreignlanguage{russian}{леди баг и супер-кот},ladybug and super cat,\foreignlanguage{russian}{леди баг и супер-кот},\foreignlanguage{russian}{ледибаг жана супер мышык},
\foreignlanguage{russian}{снежная королева},The Snow Queen,\foreignlanguage{russian}{Снежная королева},\foreignlanguage{russian}{Кар ханышасы},
\foreignlanguage{russian}{алиса в стране чудес},Alice in Wonderland,\foreignlanguage{russian}{Алиса в стране чудес},\foreignlanguage{russian}{Алиса кереметтер өлкөсүндө},
\foreignlanguage{russian}{баба яга},baba yaga,\foreignlanguage{russian}{баба яга},\foreignlanguage{russian}{баба яга},\checkmark
\foreignlanguage{russian}{детские сказки},children's fairy tales,\foreignlanguage{russian}{детские сказки},\foreignlanguage{russian}{балдар жомоктору},
\end{filecontents*}
\csvreader[
    longtable={*4{>{\centering\arraybackslash}p{3cm}}{>{\centering\arraybackslash}p{1cm}}}, 
    table head={\caption{Child search terms compiled in the course of the study. We collect the original search terms from Google Trends. Research assistants translated each term into English, Russian, and Kyrgyz, then indicated whether the term is bivalent (i.e., same translation in Russian and Kyrgyz). English translations were not tested, but are provided for readability.}\\\toprule
        \textbf{Original Child Search Term} & \textbf{English Translation} (Not Tested; Provided for Readability) & \textbf{Russian Translation} & \textbf{Kyrgyz Translation} & \textbf{Bivalent} \\         \midrule
        \endfirsthead
        \toprule
 \textbf{Original Child Search Term} & \textbf{English Translation} & \textbf{Russian Translation} & \textbf{Kyrgyz Translation} & \textbf{Bivalent}\\        \midrule
        \endhead},
    late after line=\\\midrule,
    late after last line=\\\bottomrule,
    ]{appendix_search_child_terms.csv}{1=\og,2=\eng,3=\rus,4=\kg,5=\bi }{\og & \eng  & \rus & \kg & \bi} 

\section{Adult Search Terms}
\label{appendix:adultterms}

  \begin{filecontents*}{appendix_search_adult_terms.csv}
Original Search Term,English Translation (Not Tested; Provided for Readability),Russian Translation,Kyrgyz Translation,Bivalent
\foreignlanguage{russian}{кино},movie,\foreignlanguage{russian}{кино},\foreignlanguage{russian}{кино},\checkmark
\foreignlanguage{russian}{песня},song,\foreignlanguage{russian}{песня},\foreignlanguage{russian}{ыр},
\foreignlanguage{russian}{трактор},tractor,\foreignlanguage{russian}{трактор},\foreignlanguage{russian}{трактор},\checkmark
\foreignlanguage{russian}{ырлар},songs,\foreignlanguage{russian}{песни},\foreignlanguage{russian}{ырлар},
\foreignlanguage{russian}{караоке},karaoke,\foreignlanguage{russian}{караоке},\foreignlanguage{russian}{караоке},\checkmark
\foreignlanguage{russian}{папалям папалям},across the fields across the fields,\foreignlanguage{russian}{по полям по полям},\foreignlanguage{russian}{талааларда талааларда},
\foreignlanguage{russian}{узбек кино},Uzbek film,\foreignlanguage{russian}{Узбекский фильм},\foreignlanguage{russian}{Өзбек кино},
\foreignlanguage{russian}{кыргыз кино},Kyrgyz film,\foreignlanguage{russian}{Кыргызский фильм},\foreignlanguage{russian}{Кыргыз кино},
\foreignlanguage{russian}{папа},dad,\foreignlanguage{russian}{папа},\foreignlanguage{russian}{ата},
\foreignlanguage{russian}{фильмы},films,\foreignlanguage{russian}{фильмы},\foreignlanguage{russian}{тасмалар},
\foreignlanguage{russian}{кослайк},Let's connect,\foreignlanguage{russian}{Давайте соединяться},\foreignlanguage{russian}{Кошулалы},
\foreignlanguage{russian}{мультфильм},cartoon,\foreignlanguage{russian}{мультфильм},\foreignlanguage{russian}{мультфильм},\checkmark
\foreignlanguage{russian}{машина},car,\foreignlanguage{russian}{машина},\foreignlanguage{russian}{машина},\checkmark
\foreignlanguage{russian}{мультик мультик},cartoon cartoon,\foreignlanguage{russian}{мультик мультик},\foreignlanguage{russian}{мультик мультик},\checkmark
\foreignlanguage{russian}{маша медведь},Masha Bear,\foreignlanguage{russian}{Маша Медведь},\foreignlanguage{russian}{Маша аюу},
\foreignlanguage{russian}{жанылыктар},news,\foreignlanguage{russian}{новости},\foreignlanguage{russian}{жаңылыктар},
\foreignlanguage{russian}{кыргызча кино},movie in Kyrgyz ,\foreignlanguage{russian}{кино на кыргызском},\foreignlanguage{russian}{кыргызча кино},
\foreignlanguage{russian}{кыргызча ырлар},songs in Kyrgyz,\foreignlanguage{russian}{песни на кыргызском },\foreignlanguage{russian}{кыргызча ырлар},
\foreignlanguage{russian}{музыка музыка},music music,\foreignlanguage{russian}{музыка музыка},\foreignlanguage{russian}{музыка музыка},\checkmark
\foreignlanguage{russian}{танцуй если знаешь этот тренд},dance if you know this trend,\foreignlanguage{russian}{танцуй если знаешь этот тренд},\foreignlanguage{russian}{бул трендди билсең бийле},
\foreignlanguage{russian}{бекжан темирхан},Bekjan Temirhan,\foreignlanguage{russian}{Бекжан Темирхан},\foreignlanguage{russian}{Бекжан Темирхан},\checkmark
\foreignlanguage{russian}{тик ток},tik tok,\foreignlanguage{russian}{тик ток},\foreignlanguage{russian}{тик ток},\checkmark
\foreignlanguage{russian}{настя папа},Nastya dad,\foreignlanguage{russian}{Настя папа},\foreignlanguage{russian}{Настя ата},
\foreignlanguage{russian}{бир болайык},let's be one,\foreignlanguage{russian}{давай будем одним},\foreignlanguage{russian}{бир бололу },
\foreignlanguage{russian}{музыка 2022},music 2022,\foreignlanguage{russian}{музыка 2022},\foreignlanguage{russian}{музыка 2022},\checkmark
\foreignlanguage{russian}{настя и папа},Nastya and dad,\foreignlanguage{russian}{Настя и папа},\foreignlanguage{russian}{Настя жана ата},
\foreignlanguage{russian}{узбек кино 2022},uzbek movies 2022 ,\foreignlanguage{russian}{узбекское кино 2022},\foreignlanguage{russian}{өзбек кино 2022},
\foreignlanguage{russian}{сиреноголовый},siren-headed,\foreignlanguage{russian}{сиреноголовый},\foreignlanguage{russian}{сирена баштуу},
\foreignlanguage{russian}{кошунаны тандабайт},don't choose a neighbor,\foreignlanguage{russian}{соседа не выбирают},\foreignlanguage{russian}{кошунаны тандабайт},
\foreignlanguage{russian}{фф},ff,\foreignlanguage{russian}{фф},\foreignlanguage{russian}{фф},\checkmark
\foreignlanguage{russian}{эрлан андашев},erlan andashev,\foreignlanguage{russian}{Эрлан Андашев},\foreignlanguage{russian}{Эрлан Андашев},\checkmark
\foreignlanguage{russian}{tiktok},tiktok,\foreignlanguage{russian}{тикток},\foreignlanguage{russian}{тикток},\checkmark
\foreignlanguage{russian}{кампус},campus ,\foreignlanguage{russian}{кампус},\foreignlanguage{russian}{кампус},\checkmark
\foreignlanguage{russian}{улукманапо},ulukmanapo,\foreignlanguage{russian}{улукманапо},\foreignlanguage{russian}{улукманапо},\checkmark
\foreignlanguage{russian}{инстасамка},instasamka,\foreignlanguage{russian}{инстасамка},\foreignlanguage{russian}{инстасамка},\checkmark
\foreignlanguage{russian}{фри фаер},free fire ,\foreignlanguage{russian}{фри файр},\foreignlanguage{russian}{фри файр},\checkmark
\foreignlanguage{russian}{кыргыз кино 2022},kyrgyz movies 2022 ,\foreignlanguage{russian}{кыргызское кино 2022},\foreignlanguage{russian}{кыргыз кино 2022},
\foreignlanguage{russian}{моя анимированная история},my animated story,\foreignlanguage{russian}{моя анимированная история},\foreignlanguage{russian}{менин анимацияланган окуям},
\foreignlanguage{russian}{фильмы 2022},movies 2022,\foreignlanguage{russian}{фильмы 2022},\foreignlanguage{russian}{фильмы 2022},\checkmark
\foreignlanguage{russian}{джел},jel,\foreignlanguage{russian}{джел},\foreignlanguage{russian}{джел},\checkmark
\foreignlanguage{russian}{танцуй если знаешь этот тренд 2022},dance if you know this trend 2022,\foreignlanguage{russian}{танцуй если знаешь этот тренд 2022},\foreignlanguage{russian}{бул трендди билсең бийле 2022},
\foreignlanguage{russian}{акжол махмудов роман власов},Akzhol Mahmudov Roman Vlasov,\foreignlanguage{russian}{Акжол Махмудов Роман Власов},\foreignlanguage{russian}{Акжол Махмудов Роман Власов},\checkmark
\foreignlanguage{russian}{парад победы 2023},Victory Parade 2023,\foreignlanguage{russian}{Парад Победы 2023},\foreignlanguage{russian}{Жеңиш парады 2023},
\foreignlanguage{russian}{акжол махмудов роман власов прямой эфир},Akzhol Mahmudov Roman Vlasov live broadcast,\foreignlanguage{russian}{Акжол Махмудов Роман Власов прямой эфир},\foreignlanguage{russian}{Акжол Махмудов Роман Власов түз эфир},
\foreignlanguage{russian}{ктрк спорт прямой эфир},ktrk sport live,\foreignlanguage{russian}{ктрк спорт прямой эфир},\foreignlanguage{russian}{ктрк спорт түз эфир},
\foreignlanguage{russian}{табибим мухаммадзиё мп3 скачать},Tabibim Mukhammadziyo mp3 download,\foreignlanguage{russian}{Табибим Мухаммадзиё мп3 скачать},\foreignlanguage{russian}{Табибим Мухаммадзиё мп3 көчүрүү},
\foreignlanguage{russian}{катюша},Katyusha,\foreignlanguage{russian}{Катюша},\foreignlanguage{russian}{Катюша},\checkmark
\foreignlanguage{russian}{кок бору},Kok boru,\foreignlanguage{russian}{кок бору},\foreignlanguage{russian}{көк бөрү},
\foreignlanguage{russian}{егорик},egorik,\foreignlanguage{russian}{егорик},\foreignlanguage{russian}{егорик},\checkmark
\foreignlanguage{russian}{музыка для души},music for the soul,\foreignlanguage{russian}{музыка для души},\foreignlanguage{russian}{жан үчүн музыка},
\foreignlanguage{russian}{азиз яр},aziz yar,\foreignlanguage{russian}{азиз яр},\foreignlanguage{russian}{азиз яр},\checkmark
\foreignlanguage{russian}{вальс},waltz,\foreignlanguage{russian}{вальс},\foreignlanguage{russian}{вальс},\checkmark
\foreignlanguage{russian}{yalı çapkını},Yalı Çapkını,\foreignlanguage{russian}{Ялы Чапкини},\foreignlanguage{russian}{Ялы Чапкини},\checkmark
\foreignlanguage{russian}{кослайк прямой эфир},Kos Like live,\foreignlanguage{russian}{кослайк прямой эфир},\foreignlanguage{russian}{кослайк түз эфир},
\foreignlanguage{russian}{нурайым акылбекова},Nurayym Akylbekova,\foreignlanguage{russian}{Нурайым Акылбекова},\foreignlanguage{russian}{Нурайым Акылбекова},\checkmark
\foreignlanguage{russian}{жаны ырлар 2023},New songs 2023,\foreignlanguage{russian}{новые песни 2023},\foreignlanguage{russian}{жаңы ырлар 2023},
\foreignlanguage{russian}{гулбадан},Gulbadan,\foreignlanguage{russian}{Гулбадан},\foreignlanguage{russian}{Гулбадан},\checkmark
\foreignlanguage{russian}{флешмоб},flashmob,\foreignlanguage{russian}{флешмоб},\foreignlanguage{russian}{флешмоб},\checkmark
\foreignlanguage{russian}{тойские песни},wedding songs ,\foreignlanguage{russian}{тойские песни},\foreignlanguage{russian}{той ырлары},
\foreignlanguage{russian}{караоке кыргызча},karoke in Kyrgyz,\foreignlanguage{russian}{караоке на кыргызском},\foreignlanguage{russian}{караоке кыргызча},
\foreignlanguage{russian}{4 girl team},4 girl team,\foreignlanguage{russian}{команда из 4 девушек},\foreignlanguage{russian}{4 кыздан турган команда},
\foreignlanguage{russian}{qoslike},Kos Like ,\foreignlanguage{russian}{кослайк},\foreignlanguage{russian}{кослайк},\checkmark
\foreignlanguage{russian}{агай},Male teacher/mentor ,\foreignlanguage{russian}{Преподаватель (мужчина)},\foreignlanguage{russian}{aгай},
\foreignlanguage{russian}{агай 1 серия},Male teacher/mentor 1st episode ,\foreignlanguage{russian}{Преподаватель (мужчина 1-серия)},\foreignlanguage{russian}{агай 1 серия},
\foreignlanguage{russian}{миледи},milady,\foreignlanguage{russian}{миледи},\foreignlanguage{russian}{айым},
\foreignlanguage{russian}{агай 8 серия},Male teacher/mentor 8th episode ,\foreignlanguage{russian}{Преподаватель (мужчина 8-серия)},\foreignlanguage{russian}{агай 8 серия},
\foreignlanguage{russian}{агай 3 серия},Male teacher/mentor 3rd episode ,\foreignlanguage{russian}{Преподаватель (мужчина 3-серия)},\foreignlanguage{russian}{агай 3 серия},
\foreignlanguage{russian}{зари},dawn,\foreignlanguage{russian}{зари},\foreignlanguage{russian}{таң},
\foreignlanguage{russian}{агай 2 серия},Male teacher/mentor 2nd episode ,\foreignlanguage{russian}{Преподаватель (мужчина 2-серия)},\foreignlanguage{russian}{агай 2 серия},
\foreignlanguage{russian}{миллион 2023 концерт дастури},million 2023 Dasturi concert,\foreignlanguage{russian}{миллион 2023 концерт дастури},\foreignlanguage{russian}{миллион 2023 дастури концерти},
\foreignlanguage{russian}{алматыга бараткан кыз},girl going to Almaty,\foreignlanguage{russian}{девушка едущая в Алматы},\foreignlanguage{russian}{Алматыга бараткан кыз},
\foreignlanguage{russian}{кыргызча ырлар 2023},Kyrgyz songs 2023,\foreignlanguage{russian}{Кыргызские песни 2023},\foreignlanguage{russian}{кыргызча ырлар 2023},
\foreignlanguage{russian}{оловли юрак},Olovli Yurak,\foreignlanguage{russian}{Оловли Юрак},\foreignlanguage{russian}{Оловли Юрак},\checkmark
\foreignlanguage{russian}{эгиз тумар},double talisman,\foreignlanguage{russian}{двойной талисман},\foreignlanguage{russian}{эгиз тумар},
\foreignlanguage{russian}{музыка 2023},music 2023,\foreignlanguage{russian}{музыка 2023},\foreignlanguage{russian}{музыка 2023},\checkmark
\foreignlanguage{russian}{апа},mom,\foreignlanguage{russian}{мама},\foreignlanguage{russian}{апа},
\foreignlanguage{russian}{чбд},What Happened Next(WHN),\foreignlanguage{russian}{чбд},\foreignlanguage{russian}{чбд},\checkmark
\foreignlanguage{russian}{тихий океан},Pacific ocean,\foreignlanguage{russian}{тихий океан},\foreignlanguage{russian}{тынч океан},
\foreignlanguage{russian}{танцуй если знаешь этот тренд 2023},dance if you know this trend 2023,\foreignlanguage{russian}{танцуй если знаешь этот тренд 2023},\foreignlanguage{russian}{бул трендди билсең бийле 2023},
\foreignlanguage{russian}{жалын бий},flame dance,\foreignlanguage{russian}{танец пламени},\foreignlanguage{russian}{жалын бий},
\foreignlanguage{russian}{улукбек карыбек уулу},Ulukbek Karybek uulu,\foreignlanguage{russian}{Улукбек Карыбек уулу},\foreignlanguage{russian}{Улукбек Карыбек уулу},\checkmark
\foreignlanguage{russian}{зомбеты},Zombets,\foreignlanguage{russian}{зомбеты},\foreignlanguage{russian}{зомбеты},\checkmark
\foreignlanguage{russian}{men inaya di},Men Inaya Di,\foreignlanguage{russian}{Мен Инайа Ди},\foreignlanguage{russian}{Мен Инайа Ди},\checkmark
\foreignlanguage{russian}{биржан байбол},Birzhan Baibol,\foreignlanguage{russian}{Биржан Байбол},\foreignlanguage{russian}{Биржан Байбол},\checkmark
\foreignlanguage{russian}{мультихайп},multihype,\foreignlanguage{russian}{мультихайп},\foreignlanguage{russian}{мультихайп},\checkmark
\end{filecontents*}
\csvreader[
    longtable={*4{>{\centering\arraybackslash}p{3cm}}{>{\centering\arraybackslash}p{1cm}}}, 
    table head={\caption{Adult search terms compiled in the course of the study. We collect the original search terms from Google Trends. Research assistants translated each term into English, Russian, and Kyrgyz, then indicated whether the term is bivalent (i.e., same translation in Russian and Kyrgyz). English translations were not tested, but are provided for readability.}\\\toprule
        \textbf{Original Adult Search Term} & \textbf{English Translation} (Not Tested; Provided for Readability) & \textbf{Russian Translation} & \textbf{Kyrgyz Translation} & \textbf{Bivalent} \\         \midrule
        \endfirsthead
        \toprule
 \textbf{Original Adult Search Term} & \textbf{English Translation} & \textbf{Russian Translation} & \textbf{Kyrgyz Translation} & \textbf{Bivalent}\\        \midrule
        \endhead},
    late after line=\\\midrule,
    late after last line=\\\bottomrule,
    ]{appendix_search_adult_terms.csv}{1=\og,2=\eng,3=\rus,4=\kg,5=\bi }{\og & \eng  & \rus & \kg & \bi} 

\section{Evaluation of Kyrgyz Typing Practice}

\begin{figure}[h]
  \centering
  \includegraphics[width=.8\linewidth]{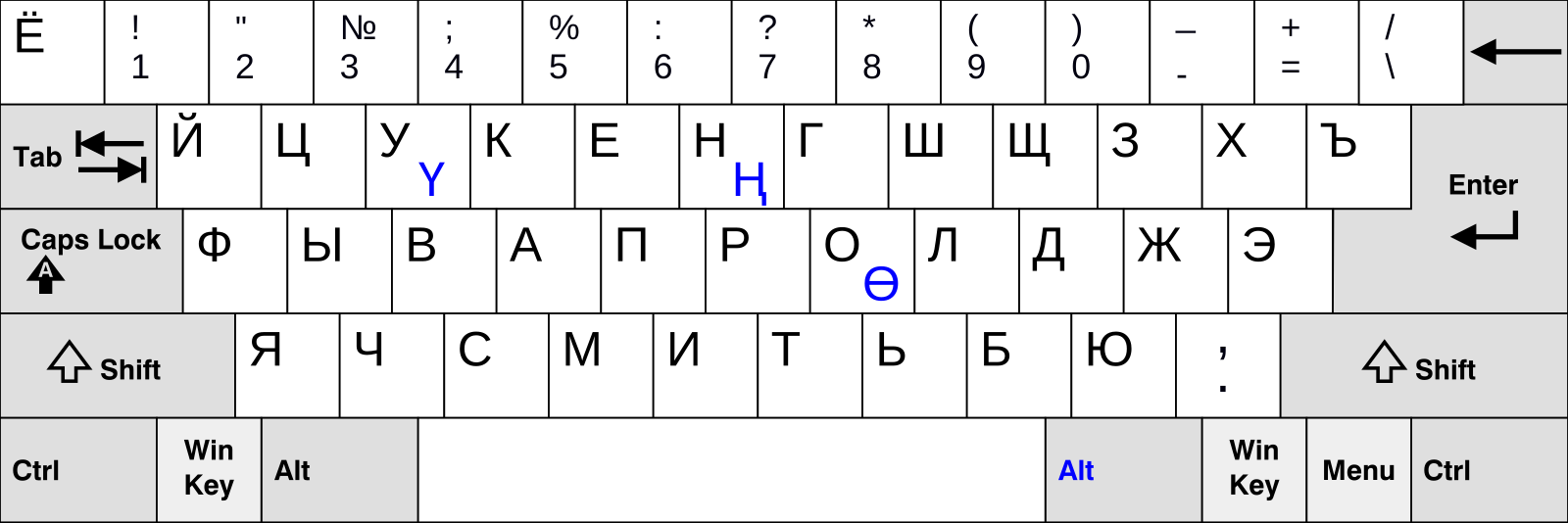}
  \caption{Kyrgyz keyboard layout. The three non-Russian Kyrgyz letters, shown in blue, can only be accessed by using a shortcut. (Source: Wikimedia Commons)}
  \Description{Layout of the Kyrygz keyboard, with non-Russian Kyrgyz characters in blue, sharing a key with a related Russian character.}
	\label{fig:keyboard}
\end{figure}

In our second experiment, involving the search bots, we additionally represented the terms according to the popular typing practice of replacing Kyrgyz letters with convenient Russian Cyrillic substitutions. We observed that the search terms from our ethnographic data and Google Trends were not formatted using the standard Kyrgyz alphabet. The Kyrgyz alphabet includes all the Cyrillic Russian letters along with \foreignlanguage{russian}{ң}, \foreignlanguage{russian}{ө}, and \foreignlanguage{russian}{ү}. However, the keyboard layout in Kyrgyzstan is identical to the Cyrillic keyboard used in Russia. Typing the letters \foreignlanguage{russian}{ң}, \foreignlanguage{russian}{ө}, and \foreignlanguage{russian}{ү} requires users to touch and hold a key on a mobile keyboard or use a keyboard shortcut on a physical keyboard (see Figure \ref{fig:keyboard}). As a result,  much typed Kyrgyz omits these non-Russian letters. For instance, the word for ``blue'' in Kyrgyz is ``\foreignlanguage{russian}{көк},'' but is often be typed as ``\foreignlanguage{russian}{кок},'' as one must either touch and hold the ``\foreignlanguage{russian}{о}'' key or use the shortcut ``Alt-\foreignlanguage{russian}{о}'' to access the Kyrgyz letter ``\foreignlanguage{russian}{ө}.'' Hence, we represented the search terms in both formal Kyrgyz and the commonly typed ``Keyboard Kyrgyz.'' This approach allows us to test the effect of the widespread practice of typing Kyrgyz using only the letters of the Cyrillic Russian alphabet. We probe the Keyboard Kyrgyz typing practice for 25 Kyrgyz terms that contained a non-Russian letter (\foreignlanguage{russian}{ң}, \foreignlanguage{russian}{ө}, or \foreignlanguage{russian}{ү}), which are listed in Appendix Section \ref{keyboardkgterms}.

\subsection{The popular Keyboard Kyrgyz typing practice produces fewer Kyrgyz and more Russian results than typing formal Kyrgyz.}

We compare terms that contain a Kyrgyz letter (e.g. ``\foreignlanguage{russian}{\textbf{ү}ч мышык}''/three cats) with their Cyrillic Russian substitutions (e.g., ``\foreignlanguage{russian}{\textbf{у}ч мышык}''). For each pair, we record the difference between the number of recommendations returned in Russian and in Kyrgyz (e.g., the Keyboard Kyrgyz term ``\foreignlanguage{russian}{\textbf{у}ч мышык}'' returned 5 more Russian and 3 fewer Kyrgyz results than the formal Kyrgyz ``\foreignlanguage{russian}{\textbf{ү}ч мышык}''). A one-sample t-test shows that the Keyboard Kyrgyz writing system results in significantly more video recommendations in Russian ($p = 0.039$) and significantly fewer in Kyrgyz ($p = 0.019$) compared to the formal Kyrgyz writing system. So, we reject the null hypothesis that there were no differences in the count of recommended videos in Russian and Kyrgyz for formal Kyrgyz and Keyboard Kyrgyz terms (see Table \ref{tab:keyboard}). These results suggest that the common practice for writing Kyrgyz online is less recognizable as Kyrgyz by the YouTube algorithm compared to the formal Kyrgyz written form. To better align recommendations with caregiver preferences, we recommend the end-user strategy of typing Kyrgyz with the full set of letters rather than using the convenient Russian Cyrillic replacements.

\begin{table}[]
\caption{Common Keyboard Kyrgyz typing practice leads to more Russian and fewer Kyrgyz videos in the top 10 search results. One-sample t-test, where the null hypothesis is that there is no difference in the count of languages recommended for Kyrgyz and Keyboard Kyrgyz terms (* = $p \textless 0.05)$.}
\begin{tabular}{llll}
 & \textbf{Mean (95\% Confidence Interval)} & \textbf{Effect Size} & \textbf{Significance} \\ \hline
\textit{Difference in Russian Results} & 1.12 ($\pm1.06$) & Cohen's d$=2.18$ & $p=0.039$* \\ \hline
\textit{Difference in Kyrgyz Results} & -1.00 ($\pm0.82$) & Cohen's d=-2.5 & $p=0.019$*
\end{tabular}%
\label{tab:keyboard}
\end{table}

\subsection{Keyboard Kyrgyz Terms}
\label{keyboardkgterms}

 \begin{filecontents*}{appendix_search_kyrgyz_letter_terms.csv}
English Translation (Not Tested; Provided for Readability),Kyrgyz Translation,Keyboard Kyrgyz
Uzbek film,\foreignlanguage{russian}{Өзбек кино},\foreignlanguage{russian}{озбек кино}
news,\foreignlanguage{russian}{жаңылыктар},\foreignlanguage{russian}{жанылыктар}
dance if you know this trend,\foreignlanguage{russian}{бул трендди билсең бийле},\foreignlanguage{russian}{бул тренди билсен бийле}
uzbek movies 2022 ,\foreignlanguage{russian}{өзбек кино 2022},\foreignlanguage{russian}{озбек кино 2022}
dance if you know this trend 2022,\foreignlanguage{russian}{бул трендди билсең бийле 2022},\foreignlanguage{russian}{бул тренди билсен бийле 2022}
Victory Parade 2023,\foreignlanguage{russian}{Жеңиш парады 2023},\foreignlanguage{russian}{жениш парады 2023}
Akzhol Mahmudov Roman Vlasov live broadcast,\foreignlanguage{russian}{Акжол Махмудов Роман Власов түз эфир},\foreignlanguage{russian}{Акжол Махмудов Роман Власов туз эфир}
ktrk sport live,\foreignlanguage{russian}{ктрк спорт түз эфир},\foreignlanguage{russian}{ктрк спорт туз эфир}
Tabibim Mukhammadziyo mp3 download,\foreignlanguage{russian}{Табибим Мухаммадзиё мп3 көчүрүү},\foreignlanguage{russian}{Табибим Мухаммадзиё мп3 кочуруу}
Kok boru,\foreignlanguage{russian}{көк бөрү},\foreignlanguage{russian}{кок бору}
music for the soul,\foreignlanguage{russian}{жан үчүн музыка},\foreignlanguage{russian}{жан учун музыка}
Kos Like live,\foreignlanguage{russian}{кослайк түз эфир},\foreignlanguage{russian}{кослайк туз эфир}
New songs 2023,\foreignlanguage{russian}{жаңы ырлар 2023},\foreignlanguage{russian}{жаны ырлар 2023}
dawn,\foreignlanguage{russian}{таң},\foreignlanguage{russian}{тан}
dance if you know this trend 2023,\foreignlanguage{russian}{бул трендди билсең бийле 2023},\foreignlanguage{russian}{бул тренди билсен бийле 2023}
bun,\foreignlanguage{russian}{көмөч нан},\foreignlanguage{russian}{комоч нан}
cartoons for boys,\foreignlanguage{russian}{балдар үчүн мультфильмдер},\foreignlanguage{russian}{балдар учун мультфильмдер}
cartoons for kids,\foreignlanguage{russian}{балдар үчүн мультфильмдер},\foreignlanguage{russian}{балдар учун мультфильмдер}
Puss in Boots (lit. Cat in Boots),\foreignlanguage{russian}{Өтүкчөн мышык},\foreignlanguage{russian}{Отукчон мышык}
Puss in Boots 2 (lit. Cat in Boots 2),\foreignlanguage{russian}{Өтүкчөн мышык 2},\foreignlanguage{russian}{Отукчон мышык 2 }
bun fairy tale,\foreignlanguage{russian}{көмөч нан жомок},\foreignlanguage{russian}{комоч нан жомок}
formidable family,\foreignlanguage{russian}{каардуу үй-бүлө},\foreignlanguage{russian}{карду уй-було}
Alice in Wonderland,\foreignlanguage{russian}{Алиса кереметтер өлкөсүндө},\foreignlanguage{russian}{Алиса кереметтер олкосундо}
\end{filecontents*}
\csvreader[
    longtable={*3{>{\centering\arraybackslash}p{4.33cm}}}, 
    table head={\caption{Search terms used to test the influence of the common Kyrgyz typing practice that substitutes Kyrgyz letters (``\foreignlanguage{russian}{ң},'' ``\foreignlanguage{russian}{ө},'' and ``\foreignlanguage{russian}{ү}'') with their Cyrillic Russian alphabet counterparts (``\foreignlanguage{russian}{н},'' ``\foreignlanguage{russian}{о},'' and ``\foreignlanguage{russian}{у}''). Research assistants collected all terms that contain Kyrgyz letters ``\foreignlanguage{russian}{ң},'' ``\foreignlanguage{russian}{ө},'' or ``\foreignlanguage{russian}{ү}'', and retyped them with their Cyrillic Russian replacements to generate the Keyboard Kyrgyz terms. English translations were not tested, but are provided for readability.}\\\toprule
        \textbf{English Translation} (Not Tested; Provided for Readability) & \textbf{Kyrgyz Translation} & \textbf{Keyboard Kyrgyz} \\         \midrule
        \endfirsthead
        \toprule
 \textbf{English Translation} & \textbf{Kyrgyz Translation} & \textbf{Keyboard Kyrgyz} \\        \midrule
        \endhead},
    late after line=\\\midrule,
    late after last line=\\\bottomrule,
    ]{appendix_search_kyrgyz_letter_terms.csv}{1=\eng,2=\kg,3=\keykg }{\eng & \kg & \keykg}

\end{document}